\documentclass[journal]{IEEEtran}
\IEEEoverridecommandlockouts
\usepackage{cite}
\usepackage{amsmath,amssymb,amsfonts}
\usepackage[ruled,vlined,linesnumbered]{algorithm2e}
\usepackage{graphicx}
\usepackage{textcomp}
\usepackage{tikz}
\usepackage{xcolor,colortbl}

\usepackage[switch]{lineno}

\usepackage[normalem]{ulem}
\newcommand{\stkout}[1]{\ifmmode\text{\sout{\ensuremath{#1}}}\else\sout{#1}\fi}

\def \nobreakseq {\nobreak \hskip 0pt \hbox}
\def\BibTeX{{\rm B\kern-.05em{\sc i\kern-.025em b}\kern-.08em
    T\kern-.1667em\lower.7ex\hbox{E}\kern-.125emX}}

\usepackage{stfloats}
\usepackage{bm}

\usepackage{amsthm}
\newtheorem{prop}{Proposition}
\newtheorem{SampleEnv}{Example}

\tikzset{every picture/.style={line width=0.75pt}} 

\DeclareMathOperator*{\argmax}{arg\,max}

\newcommand{\SNR}{\mathrm{SNR}\xspace}
\newcommand{\SF}{\mathrm{SF}\xspace}
\newcommand{\TDEL}{\mathrm{TDEL}\xspace}
\newcommand{\coh}{\mathrm{coh}\xspace}
\newcommand{\nocoh}{\mathrm{non\text{-}coh}\xspace}
\newcommand{\mf}{\mathrm{mf}\xspace}
\newcommand{\idealmf}{\mathrm{ideal}\text{-}\mathrm{mf}\xspace}
\newcommand{\avg}{\mathrm{avg}\xspace}
\newcommand{\norm}{\mathrm{norm}\xspace}

\newcommand\T{\rule{0pt}{2.2ex}}  

\let\Gamma\varGamma

\begin{document}

\title{Simple and Efficient LoRa Receiver Scheme for Multi-Path Channel}


\author{Clément~Demeslay,~\IEEEmembership{Student Member,~IEEE,}
        Philippe~Rostaing,~\IEEEmembership{Member,~IEEE,}
        and~Roland~Gautier,~\IEEEmembership{Member,~IEEE}
\IEEEcompsocitemizethanks{\IEEEcompsocthanksitem C. Demeslay, P. Rostaing and R. Gautier are with CNRS UMR 6285, Lab-STICC, from University of Brest, CS 93837, 6 avenue Le Gorgeu, 29238 Brest Cedex 3, France.\protect\\
E-mail: clement.demeslay@univ-brest.fr, philippe.rostaing@univ-brest.fr, roland.gautier@univ-brest.fr}
}

\makeatletter\def\IEEElabelanchoreqn#1{\bgroup\def\@currentlabel{\p@equation\theequation}\relax\def\@currentHref{\@IEEEtheHrefequation}\label{#1}\relax\Hy@raisedlink{\hyper@anchorstart{\@currentHref}}\relax\Hy@raisedlink{\hyper@anchorend}\egroup}\makeatother\newcommand{\subnumberinglabel}[1]{\IEEEyesnumber\IEEEyessubnumber*\IEEElabelanchoreqn{#1}}


\maketitle

\begin{abstract}
This paper presents a novel LoRa (Long Range) receiver operating in frequency selective Multi-Path Channel (MPC).
The dechirped received LoRa wave-forms under MPC allows us to derive a simple and efficient LoRa receiver scheme by using a MF (Matched Filter)
approach that aims to maximize the SNR (Signal-to-Noise Ratio) at the symbol index frequency of the DFT output.
We show that the MF receiver can be seen as RAKE structure where interference peaks related to multi-path, exhibited at DFT output, are recombined in a constructive way. 
Detection performance is driven by channel energy and the benefit of this novel MF/RAKE receiver over original coherent and non-coherent receiver appears only for MPC that exhibits significant paths energy.
These two MF and RAKE receivers have however different implementation complexities that are studied in details.
We provide in that sense recommendations on which receiver variant to use for real operations, depending on complexity constraints.
Finally, the proposed MF/RAKE receiver outperforms previous results on TDEL (Time Delay Estimation LoRa) receiver over MPC, especially at low SNR and higher LoRa Spreading Factor (SF) parameter, at the cost of higher but reasonable complexity.
\end{abstract}


\section{Introduction}

The Internet of Things (IoT) is experiencing striking growth enabling many more devices to communicate with each other and allowing many futuristic scenarios to be a reality such as smart cities or Industry 4.0.
\cite{IoTDevicesNumber} affirms that the expected number of active IoT devices will rapidly grow to reach almost 31 billion until 2025 \cite{IoTDevicesNumber}.
Since the past few years, many IoT technologies were developed relying on existing infrastructures such as Narrow-Band IoT (NB-IoT) or LTE-Machine (LTE-M), or more dedicated networks leveraging unlicensed bands such as SigFox, Ingenu, Weightless-P or Long Range (LoRa) \cite{goursaud}.
LoRa is nowadays a front runner of Low-Power Wide Area Networks (LP-WAN) solution and holds a lot of attention by the scientific research community.
We will focus on LoRa in this paper.
Due to its patented nature, initial research was mainly based on retro-engineering of existing LoRa transceivers \cite{knight_gnuradio}. 
One of the first paper to provide a rigorous mathematical representation of LoRa signals and its demodulation scheme was achieved in \cite{vangelista}.
Since then, further researches were conducted focusing on specific issues such as network capacity enhancements \cite{joerg3}, channel coding improvements \cite{joerg4,imt}, temporal/frequency synchronization schemes \cite{xhonneux_2021}, LoRa demodulation with LoRa interference \cite{laporte,xhonneux_interf} or LoRa implementation on real world equipment such as Universal Software Radio Equipment (USRP) \cite{tapparel}.

More specifically, recent researches addressed LoRa multi-path channel (MPC) impact and equalization issues \cite{bapathu,demeslay,guo}.
The authors in \cite{bapathu} investigated the impact of a rapidly-varying channel on LoRa Frame Error Rate (FER).
They came to the conclusion that the FER performance depends on a Spreading Factor (SF) parameter-\textcolor{black}{frame length} trade-off.
In their analytical study, \textcolor{black}{the authors in \cite{demeslay}} highlighted that although LoRa wave-forms experience small Inter Symbol Interference (ISI) most of the time in practice, the basic LoRa demodulator is very sensitive to one or several significant echoes.  \textcolor{black}{In other words, uncoded SER performance degradation is neglected only for attenuation of echoes, relative to the peak-value of the direct path, that are greater than 14~dB.}
To overcome the issue of higher significant echoes, the authors in \cite{guo} proposed an enhanced {non-coherent} LoRa receiver, exhibiting good performance.
However, due to the non-coherent nature of this receiver, the LoRa Symbol Error Rate (SER) performance may be further improved by considering a coherent approach.
This was stated in \cite{afisiadis3} for LoRa interference scenario only, but the authors in \cite{demeslay} also highlighted that MPC effect on LoRa signals is very similar to the LoRa interference impact with the same SF.
That is, the advantages of the coherent approach may also be valid for MPC.
Then, we propose in this article a custom coherent LoRa receiver based on Matched Filtering (MF) or RAKE that leverages energy of MPC to improve the SER performance.
To the best of our knowledge, a such coherent approach for MPC has not been addressed yet in the literature.

As a first approximation, the discrete-time channel model is considered where channel path delays are supposed to be multiple of the sampling rate.

The key contributions of the paper are as follows:
\begin{itemize}
    \item A novel MF or equivalent RAKE detector is derived and outperforms in terms of SER \textit{i)} the original coherent and non-coherent LoRa detectors, and \textit{ii)} the designed detector for MPC \cite{guo} that is based on cross-correlation  between  averaged  pilot symbols  and data symbols DFT’s.
    \item  Theoretical  findings are provided to quantify the gain of the proposed detector by using a simple performance indicator based on the ratio of parasitic peaks amplitude (due to multi-paths) over amplitude of the peak of interest at the DFT output of the receiver.
        \item 
    A complexity analysis is used to select an appropriate algorithm (MF vs. RAKE) by evaluating the total required complex addition/multiplication operations, and by analyzing the execution time of the compiled algorithms.
     \item 
     A variant of our detector ("candidate" approach) is derived for practical usage including channel parameters estimation and 
     SER performance-complexity trade-offs.
\end{itemize}

The remainder of the paper is organized as follows.
In Section \ref{sec:LoRaOverview}, we present the basics of LoRa modulation and continue with the derived mathematical expressions of LoRa signals impacted by the MPC in Section \ref{sec:LoRaPartMPC}.
In Section \ref{sec:RakeMF}, we introduce the MF receiver and prove that the latter is equivalent to the RAKE receiver.
We also present a variation of this receiver to be implemented in practice and study in detail complexity in terms of total 
number of
complex operations and execution time.
This receiver requires channel knowledge and we propose in Section \ref{sec:ChannelEst} a simple scheme to estimate the channel impulse response.
Simulation results are provided in Section \ref{sec:Simu} to assess our receiver and a comparison with previous research work \cite{guo} is performed.
Finally, we conclude the article in Section \ref{sec:conclu} by providing recommendations on which receiver variant to use for real world operations.
Throughout the paper, the notations reported in Table \ref{tab:notations} are used.

\begin{table}[ht]
\setlength{\tabcolsep}{2pt}
\definecolor{Gray}{gray}{0.85}\newcolumntype{a}{>{\columncolor{Gray}}l}
\begin{center}
{\scriptsize
\begin{tabular}{a|l||a|l}
\multicolumn{4}{l}{\footnotesize \textbf{Notation and symbols meaning}}\\\hline\hline
$T$ & symbol period & $n$ & frequency index \\ \hline
$F_s$ & sampling frequency & $i$ & multi-path index \\ \hline
$T_s$ & sampling period & $p$ & pilot symbol index \\ \hline
$B$ & LoRa bandwidth & $k$ & time index \\ \hline
SF & spreading factor & $u$ & candidate symbol index \\\hline
$M$ & 
\multicolumn{3}{l}{number of possible chirp waveforms per symbol: $2^{\SF}$}\T\\
\hline 
$a$ & current transmitted symbol
& $N_d$ & number of payload symbols  \\ \hline 
$a^-$ & previous symbol from $a$ &$N_p$ & number of pilot symbols \\ \hline 
$b$ & candidate symbol
& $N_f$ & 
 \hspace{-2pt}\begin{tabular}{l}number of symbols in the frame: \\[-.1em]$N_f = N_p + N_d$ \end{tabular} \\ \hline
$b_u$ & $u$-th candidate symbol&$\lambda_c$ & threshold for candidate selection  \\ \hline
$\mathcal{A}$ & candidate symbols set& $\lambda_p$ & threshold for multi-path detection \\ \hline
$N_c$ & \hspace{-2pt}\begin{tabular}{l}number of candidate symbols:\\[-.1em] \#$\mathcal{A}$ (cardinality)\end{tabular} 
& $\lambda_{\TDEL}$ & \hspace{-2pt}\begin{tabular}{l}threshold for multi-path selection\\[-.1em]
in TDEL receiver \cite{guo}\end{tabular}\\
\hline 
$x_a[k]$ & transmitted $a$-symbol waveform
&$r_a[k]$ & received $a$-symbol waveform \\\hline 
$\tilde{r}_a[k]$ & 
\multicolumn{3}{l}{received $a$-symbol down-chirp (DC) waveform: $r_a[k]x_0^*[k]$}\T\\\hline 
$\tilde{R}_a[n]$ & 
\hspace{-2pt}\begin{tabular}{l}
legacy LoRa demodulator:\\[-.1em]
$M$-size DFT of $\tilde{r}_a[k]$\T
\end{tabular}
& $K$ & number of paths of MPC\T\\
\hline 
$\alpha(i)$ & channel gain of $i$-th path 
&$k_i$ & channel delay  of the $i$-th path \\ \hline
$\tilde{\alpha}_b(i)$ & $\tilde{\alpha}_b(i)={\alpha}(i)x_b[-k_i]$ \T
&$\tilde{\alpha}(i)$ & $\tilde{\alpha}(i)=\tilde{\alpha}_0(i)$ 
\\\hline
$C_b[k]$ & \multicolumn{3}{l}{
\hspace{-2pt}\begin{tabular}{l}
channel frequency response associated with $\tilde{\alpha}_b(i)$ at frequency $k/M$:\T\\
$M$-size DFT of $\tilde{\alpha}_b(i)$
\end{tabular}
}
\T\\\hline
$\tilde{z}_{ab}[k]$& 
\multicolumn{3}{l}{MF received waveform: $C_b^*[k]\tilde{r}_a[k]$}\T \\\hline
$\tilde{Z}_{ab}[b]$ &
\multicolumn{3}{l}{
\hspace{-2pt}\begin{tabular}{l}
RAKE or MF statistic for transmitted $a$-symbol and candidate $b$-symbol:\T\\[-.1em]
$M$-size DFT of $\tilde{z}_{ab}[k]$, and frequency index $b$ is selected
\end{tabular}}\\\hline
$w[k]$&AWGN & $\tilde{w}[k]$& DC AWGN: $w[k]x_0^*[k]$\T\\\hline
$\tilde{w}_{b}[k]$& 
MF DC AWGN: $C_b^*[k]\tilde{w}[k]$
&$\tilde{W}_{b}[k]$ & $M$-size DFT of  $\tilde{w}_{b}[k]$\T
\\\hline
 $\Gamma^{\Tilde{\alpha}}_{a,b}[l]$ & 
 \multicolumn{3}{l}{cross-correlation function of $\Tilde{\alpha}_a(m)$ and $\Tilde{\alpha}_b(m)$} \T\\\hline
\end{tabular}
}
\end{center}
{\caption{List of principal notations used in the paper.}}
\label{tab:notations}
\end{table}

\section{LoRa modulation overview}\label{sec:LoRaOverview}

\subsection{LoRa wave-forms}\label{subsec:LoRaSignals}

In the literature, LoRa wave-forms are of the type of Chirp Spread Spectrum (CSS) signals.
These signals rely on complex sine waves with Instantaneous Frequency (IF) that varies linearly with time over frequency range $f \in [-B/2,B/2]$ and time range $t \in [0,T]$ ($T$ the symbol period).
This basic signal is called an up-chirp (UC) or down-chirp (DC) when IF respectively increases or decreases over time.
A LoRa waveform is a $M$-ary digital modulation, constituted of $M$ possible chirp modulations where the IF of the UC is shifted by the $M$ possible values.
The modulo operation is applied to ensure that frequency remains in the interval $[-B/2,B/2]$.
The LoRa parameters are chosen such that $BT=M$ with  $M=2^{\SF}$ and $\SF\in \{7,8,\ldots,12\}$ is called the spreading factor, which also corresponds to the number of bits for a LoRa symbol. 

In the discrete-time signal model, the chip rate ($R_c=1/T_c=M/T$) is usually used to sample the received signal, i.e., the sample period is $T_s = T_c = T/M = 1/B$.
The signal has then $M$ samples over one symbol period $T$.
Each symbol $a \in \{0,1,\ldots,M-1\}$ is mapped to an UC that is temporally shifted by $\tau_a = a T_c$ period.
We may notice that a temporal shift conducts to a change of initial IF.
This behavior is the heart of the $M$-ary chirp modulation.
A mathematical expression of LoRa wave-form sampled at $t = k T_s$ ($T_s=T_c$) has been derived in \cite{chiani}:
\begin{equation}
    x(kT_s;a) \triangleq x_a[k] = e^{2j\pi k \left( \frac{a}{M} - \frac{1}{2} + \frac{k}{2M} \right)} \quad k = 0,1,\ldots,M-1.
    \label{eq:xa}
\end{equation}

We may see that an UC is actually a LoRa wave-form with symbol index $a = 0$, written $x_0[k]$.
Its conjugate $x_0^*[k]$ is then a DC. 

\begin{figure}[htbp]
  \centering
  \includegraphics[width=0.49\textwidth]{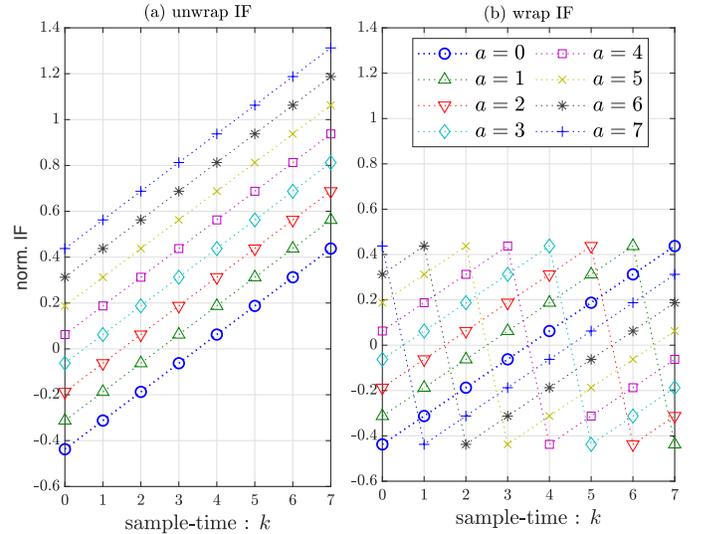}
  \caption{Normalized discrete-time IF for $M=8$, (a) unwrap IF and (b) wrap IF in $[-1/2,\,1/2[$.}
\label{fig:freqinst}
\end{figure}

By using the forward finite phase differences $\phi[k+1]-\phi[k]$  of \eqref{eq:xa}, the discrete-time IF for a given $a$-symbol is: $f_a[k]=\frac{1}{2\pi}(\phi[k+1]-\phi[k])=\frac{a+k}{M}-\frac{1}{2}+\frac{1}{2M}$.
In order to illustrate the $M$-ary chirp modulation, Fig.~\ref{fig:freqinst} plots the discrete-time IF for each transmitted symbol $a\in\{0,\ldots,7\}$ for only $\SF=3$ ($M=8$).
Note that the modulo $B$ operation is intrinsic in the discrete-time model.

In LoRa transmissions, the SNR (defined as the ratio of signal power to noise power) in the signal bandwidth is given by: $\SNR = P_s/(N_0B) = (P_s/N_0)\times(T/M) = E_s/(N_0M)$ with $N_0=E[|w[k]|^2]=\sigma^2$ and $E_s=E[[x_a[k]|^2]=1$.
The SNR per bit $E_b/N_0$ can be expressed as:
\begin{equation}
    E_b/N_0 = \SNR \times \frac{M}{\SF}.
    \label{eq:SNREbNo}
\end{equation}

\subsection{LoRa demodulation scheme}\label{subsec:LoRaDemodScheme}

Reference \cite{vangelista} derived a simple and efficient solution to demodulate LoRa signals.
In Additive White Gaussian Noise (AWGN) flat-fading channel, the demodulation process is based on the Maximum Likelihood (ML) detection scheme.

The received signal is:
\begin{equation}
    r[k] = \alpha x_a[k] + w[k]
\end{equation}
with $\alpha$ the complex gain  of the flat fading channel and $w[k]$ an independent and identical distributed (i.i.d.) complex AWGN with zero-mean and variance $\sigma^2 = E[|w[k]|^2]$.
ML detector aims to select index $n$ that maximizes the scalar product $\langle r,x_n \rangle\text{ for } n \in \{0,1,\ldots,M-1\}$ defined as:
\begin{equation}
\begin{split}
    \langle r,x_n \rangle &= \sum_{k=0}^{M-1} r[k] x_n^*[k] \\
     &=  \sum_{k=0}^{M-1} \underbrace{r[k]x_0^*[k]}_{\Tilde{r}[k]}  e^{-j2\pi\frac{n}{M}k} = \Tilde{R}[n].
\end{split}
\end{equation}

The demodulation stage proceeds with two simple operations:
\begin{itemize}
    \item multiply received wave-form by a DC $x_0^*[k]$, also called dechirping,
    \item compute $\Tilde{R}[n]$, the Discrete Fourier Transform (DFT) of $\Tilde{r}[k]$ and select the discrete frequency index $\widehat{a}$ that maximizes $\Tilde{R}[n]$.
    \label{eq:DemodLoRa}
\end{itemize}

This way, the dechirp process merges all the signal energy in a unique frequency bin $a$ and can be easily retrieved by taking the magnitude (non-coherent detection) of $\Tilde{R}[n]$.
The symbol detection is then:
\begin{equation}
    \widehat{a} = \underset{n}{\argmax} \quad |\Tilde{R}[n]|.
    \label{eq:DemodLoRaNCOH}
\end{equation}

An improvement about 1 dB \cite{imt} can be obtained with the coherent detection scheme:

\begin{equation}
    \widehat{a} = \underset{n}{\argmax} \quad \Re\{\Tilde{R}[n]\}.
    \label{eq:DemodLoRaCOH}
\end{equation}

However, a channel phase estimation and compensation must be performed before using the coherent scheme to obtain:
$r'[k]=e^{-j\phi}r[k] = |\alpha| x_a[k]+w'[k]$ with $w'[k]=e^{-j\phi}w[k]$ and $\phi=\arg(\alpha)$.

\section{Multi-path channel on LoRa signal}\label{sec:LoRaPartMPC}
\subsection{Multi-path channel model}\label{subsec:MPCModel}

We study in this section the effect of MPC on LoRa signals.
By considering the chip rate $R_c$ to sample the received signal, the equivalent discrete-time channel model is:
\begin{equation}
     c[k] = \sum_{i=0}^{K-1} \alpha(i) \delta[k - k_i]
    \label{eq:ChannelModel}
\end{equation}
with $K$ the number of paths and $\alpha(i)=|\alpha(i)|e^{j\phi(i)}$ the complex gain at path delay $\tau_i=k_iT_c$ ($k_i$ tap).

A sufficient condition to consider a channel as frequency selective is $k_i \ge 1$ ($\tau_i \ge T_c$). e.g. $B = 500$~kHz, $T_c = 1/B = 2~\mu$s.
This value is a typical path delay seen in outdoor environments (few $\mu$s usually) \cite{cost_207}.
As a symbol duration is $M \times T_c$, we suppose that the largest echo $k_{max} \ll M$.
We expect therefore to have a channel effect that introduces ISI only between the current and previous symbol over a reduced number of samples.
The basic symbol detector presented herein is very sensitive to significant paths although ISI depth is very small.
In this section, we evaluate the impact of MPC on LoRa wave-forms.
We consider a set of transmitted symbols $a_s$ ($s=0,1,\ldots,S-1$) as:
\begin{equation}
     s[k^{\prime}] = \sum_{s=0}^{S-1} x_{a_s}[k^{\prime}\!\!\mod M]
    \label{eq:sk}
\end{equation}
for $k^{\prime} = k + sM$ and $k=0,\ldots,M-1$.
The received signal is then:
\begin{equation}
     r[k^{\prime}] = \underbrace{c[k^{\prime}] * s[k^{\prime}]}_{m[k^{\prime}]} + w[k^{\prime}]
    \label{eq:yk}
\end{equation}
where the operator $*$ denotes the discrete convolution.
We note $\sigma_{\Re}^2 = \sigma_{\Im}^2 = \sigma^2/2$, the variance of real and imaginary part of $w[k^\prime]$.
The signal $m[k^{\prime}]$ is the received waveform after channel effect.

\subsection{Channel effect on LoRa wave-forms}\label{subsec:LoRaChannelEffect}

Let us denote $r_a[k]$ the received signal for detecting the current
symbol $a$ into its symbol interval for $k=0,\ldots,M-1$.
We suppose that the receiver is synchronized on the first path (i.e. $k_0=0$).

\begin{prop}
Performing the DC operation $x_0^*[k]$ to $r_a[k]$ yields:
\setlength\arraycolsep{2pt}
\begin{eqnarray}
\nonumber
\Tilde{r}_a[k] &=& x_0^*[k]r_a[k]\\
&=&\alpha(0) e^{2j\pi k \frac{a}{M}} +
  \sum_{i=1}^{K-1} \Tilde{\alpha}_{\bar{a}}(i) e^{2j\pi k
    \frac{\overline{a} - k_i}{M}} + \underbrace{x_0^*[k] w[k]}_{\Tilde{w}[k]}\nonumber\\
  \label{eq:rak2_IES}
\end{eqnarray}

\vspace{-2em}
\noindent where:
\begin{eqnarray}
 \Tilde{\alpha}_{\bar{a}}(i) &=& \alpha(i) x_{\bar{a}}[-k_i]\nonumber\\ 
 &=&
 \alpha(i) e^{-2j\pi k_i (-\frac{1}{2} - \frac{k_i}{2M})} e^{-2j\pi k_i \frac{\bar{a}}{M}}\nonumber\\
 &=&\Tilde{\alpha}(i) e^{-2j\pi k_i \frac{\bar{a}}{M}}\label{eq:alpha_i_tilde}
\end{eqnarray}
with:
\begin{eqnarray}
    \Tilde{\alpha}(i)&=&\alpha(i) e^{-2j\pi k_i (-\frac{1}{2} - \frac{k_i}{2M})}\nonumber\\
    &=&\alpha(i) x_0[-k_i]
\end{eqnarray}
and:
\begin{equation}
\bar{a} \triangleq 
\begin{cases}
  a^- & \text{ for } k = 0,\ldots,k_i-1 \hfill\text{ (previous symbol)} \\
  a & \text{ for } k = k_i,\ldots,M-1 \hfill \text{ (current symbol)}
\end{cases}
\label{eq:bar_a}
\end{equation}
Note that dechirping process does not change noise statistic.
\end{prop}

\begin{figure}[htbp]
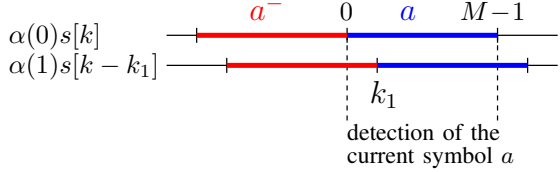

  \centering
  \input IES_symbol_LoRa.tex
  \caption{Illustration of ISI for detecting of the current symbol $a$ in case
  on two-path channel at delays $k_0=0$ (synchronized on the first path) and $k_1$.}
\label{fig:ISI}
\end{figure}

\begin{proof}
For the sake of simplicity, we first consider the two-path channel.  
The received signal is then $r[k]=\alpha(0) s[k]+\alpha(1) s[k-k_1] + w[k]$.
By focusing in the detection interval $k=0,\ldots,M-1$ of the current symbol $a$ (Fig.~\ref{fig:ISI}), the signal on the synchronized path is equal to $s[k]=x_a[k]$ and the signal on the delayed path can be expressed as:

\begin{equation}
  s[k-k_1]=\left\{
\begin{array}{ll}
x_{a^{-}}[M-k_1+k] &\text{for } k=0,\ldots,k_1-1 \\
x_{a}[k-k_1] &\text{for } k=k_1,\ldots,M-1.
\end{array}
\right.
\end{equation}

From \eqref{eq:xa} one can verify the property $x_a[M-n]=x_a[-n]$ for $n=0,1,\ldots,M-1$,
then the received signal for the detection of the current symbol $a$ could be
expressed as:

\begin{equation}
r_a[k]=\alpha(0)x_a[k]+\alpha(1)x_{\bar{a}}[k-k_1] + w[k]
\label{eq:r_a[k]}
\end{equation}
where $\bar{a}$ is defined in \eqref{eq:bar_a}.
By multiplying the DC to $r_a[k]$ we obtain after some basic manipulations:

\begin{equation}
    \begin{split}
        \Tilde{r}_a[k] = x_0^*[k] r_a[k]
        = \alpha(0)e^{j2\pi k\frac{a}{M}}+\Tilde{\alpha}_{\bar{a}}(1)e^{j2\pi n\frac{\bar{a}-k_1}{M}} + \Tilde{w}[k]
    \end{split}
\end{equation}
 with 
$\Tilde{\alpha}_{\bar{a}}(1)=\alpha(1)x_{\bar{a}}[-k_1]=\alpha(1)x_{\bar{a}}[M-k_1]$.

By applying the same development for $K>2$ paths, the general expression is
straightforward and given in \eqref{eq:rak2_IES}.
\end{proof}

As channel delay spread is small in comparison with symbol duration, we may omit interference coming from previous symbol $a^-$.
Equation~\eqref{eq:rak2_IES} can be then simplified as:
\begin{equation}
    \Tilde{r}_a[k] \approx \alpha(0) e^{2j\pi k \frac{a}{M}} + 
  \sum_{i=1}^{K-1} \Tilde{\alpha}_a(i) e^{2j\pi k \frac{a - k_i}{M}} + \Tilde{w}[k]
  \label{eq:rak2}
\end{equation}
or identically:
\begin{equation}
    \begin{split}
       \Tilde{r}_a[k] \approx C_a[k] e^{2j\pi k \frac{a}{M}} + \Tilde{w}[k]
    \end{split}
    \label{eq:ykDech2}
\end{equation}
where $C_a[k]$ can be seen as a channel coefficient, but depending on the transmit LoRa symbol $a$, and given by:
\begin{equation}
\label{eq:path}
C_a[k]=\sum_{i=0}^{K-1} \Tilde{\alpha}_a(i) e^{-2j\pi k \frac{k_i}{M}}.
\end{equation}

From (\ref{eq:alpha_i_tilde}), the channel-like path gain $\Tilde{\alpha}_a(i)$ in (\ref{eq:rak2}) or (\ref{eq:path}) is given by:
\begin{equation}
\label{eq:alphat_a}
\Tilde{\alpha}_a(i)=\Tilde{\alpha}_0(i)e^{-2j\pi k_i\frac{a}{M}}
\end{equation}
with $\Tilde{\alpha}_0(i)=\tilde{\alpha}(i)$ for $i=0,\ldots,K-1$.
Note that for the synchronized path $k_0=0$ ($i=0$), we have $\tilde{\alpha}_a(0)=\tilde{\alpha}(0)=\alpha(0)$.

\subsection{DFT of the received down-chirp LoRa signal}\label{subsec:LoRaChannelEffectDFT}

The second operation in the demodulation stage is to compute the DFT of  
$\Tilde{r}_a[k]$ and select the discrete frequency index that maximizes the DFT magnitude.
The $M$-point DFT of $\{\Tilde{r}_a[k]\}_{k=0}^{M-1}$ in (\ref{eq:rak2}) is given by:
\setlength\arraycolsep{2pt}
\begin{equation}
  \label{eq:Rak2}
  \begin{split}
  \Tilde{R}_a[n]=M\alpha(0)\delta[n-a]+M\sum_{i=1}^{K-1} \Tilde{\alpha}_a(i) \delta[n - a + k_i]\\ + \Tilde{W}[n]
\end{split}
\end{equation}
for $n=0,\ldots,M-1$,
where $\Tilde{W}[n]$ is the $M$-point DFT of $\{\tilde{w}[k]\}_{k=0}^{M-1}$ with $\Tilde{W}[n] \sim \mathcal{CN}(0,\sigma_w^2 = M \sigma^2)$.
The Kronecker function is defined as: $\delta[n] = 1$ for $n = 0\mod M$, and 0 otherwise.

\begin{figure}[ht]
  \centering
    \includegraphics[width=0.49\textwidth]{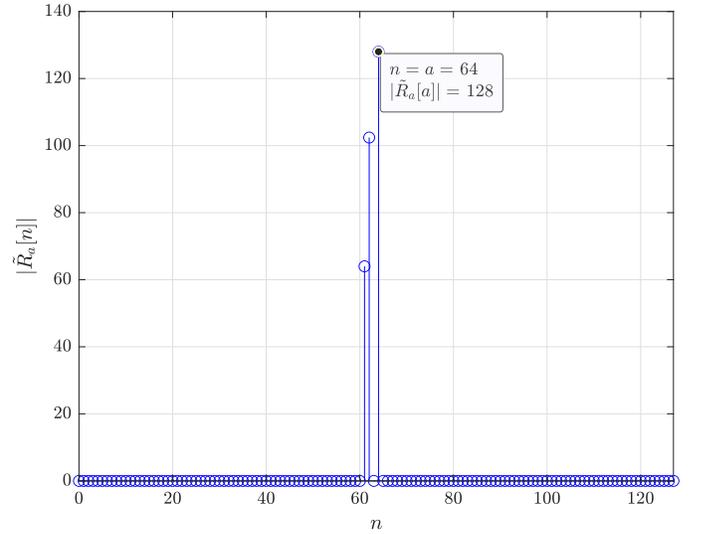}
  \caption{An illustration of the DFT magnitude of the noise-free received DC LoRa symbol with value $a=64$ passed through C1-channel, $\SF = 7$}
\label{fig:LoRa_DFT_Example}
\end{figure}

From \eqref{eq:Rak2}, we note that channel effect after demodulation stages (i.e. dechirp and DFT) consists in multiple peaks of complex amplitude $\Tilde{\alpha}_a(i)$ shifted by $k_i$ from the transmitted symbol index $a$, for $i=0,\ldots,K-1$.
When taking the DFT magnitude, the peak magnitudes are only driven by the magnitude of the path gains $|\tilde{\alpha}_a(i)|=|\alpha(i)|$.
Fig.~\ref{fig:LoRa_DFT_Example} illustrates the approximated DFT magnitude of a received DC LoRa symbol with value $a=64$ and passed through the following example channel (denoted C1-channel):
\begin{equation}
    C_1[k] = \delta[k] + 0.8 \delta[k-2] + 0.5\delta[k-3].
    \label{eq:channel_example}
\end{equation}

Note that in the case of a non-aligned MPC i.e. $k_i$ real valued, the DFT bin energy of each channel path is spread over neighbor bins.
See Fig.~3 in \cite{demeslay} for illustration.

\section{Matched Filter or RAKE receivers}\label{sec:RakeMF}

In this section, we present Matched Filter (MF) receiver which can be seen as RAKE receiver for LoRa transmission.
We suppose for now that channel path gains $\Tilde{\alpha}_0(i)=\Tilde{\alpha}(i)$ are perfectly estimated (or known) by the receiver for $i=0,\ldots,K-1$.
We will propose in the next section a method to estimate the number of paths $K$ and the associated channel path gains $\tilde{\alpha}(i)$ (at the corresponding tap $k_i$). 
In Section~\ref{sec:Simu}, performance in terms of SER is compared for both the terms $\tilde{\alpha}(i)$ known or estimated by using a preamble. 

\subsection{Ideal-MF receiver}\label{subsec:IdealMF}

A Matched-Filter approach is proposed to derive a new detection scheme only by multiplying the received DC signal $\Tilde{r}_a[k]$ in (\ref{eq:ykDech2}) by the MF coefficient $ C_a^*[k]$ to obtain:
\begin{equation}
    \begin{split}
        \Tilde{z}_{a,a}[k] = C_a^*[k] \Tilde{r}_a[k].
    \end{split}
    \label{eqMFFilter0}
\end{equation}

This MF approach is denoted ``ideal-MF" because the transmitted symbol $a$ is necessary to compute the channel coefficient $C_a[k]$, but $a$ is of course not available in practice.
The ideal-MF will lead to the best SER performance in comparison with other more practical MF approaches proposed in the following subsections, where the symbol $a$ is not supposed to be known.

By separating signal and noise terms, \eqref{eqMFFilter0} yields:
\begin{equation}
    \begin{split}
        \Tilde{z}_{a,a}[k] = |C_a[k]|^2 e^{j2\pi\frac{a}{M}k}+\Tilde{w}_a[k]
    \end{split}
    \label{eqMFFilter}
\end{equation}
where the resulting noise $\Tilde{w}_a[k]=C_a^*[k] \Tilde{w}[k]$ is zero-mean i.i.d. complex Gaussian process with time-varying noise power $E[\Tilde{w}_a[k]|^2]=\sigma^2|C_a[k]|^2$.

The  square magnitude of the channel coefficient $|C_a[k]|^2$ in (\ref{eqMFFilter}) can be expressed as:
\begin{eqnarray}
\label{eq:Ca2}
|C_a[k]|^2&=&\sum_{m,n=0}^{K-1}\Tilde{\alpha}_a(m)\Tilde{\alpha}_a^*(n)e^{-j2\pi\frac{k}{M}(k_m-k_n)}\\\label{eq:Ca3}
&=&\sum_{l=-l_{\max}}^{l_{\max}} \Gamma^{\Tilde{\alpha}}_{a,a}[l]e^{-j2\pi\frac{k}{M}l}
\end{eqnarray}
where $l_{\max}=k_{K-1}$ the tap of the last echo, and $\Gamma^{\Tilde{\alpha}}_{a,a}[l]$  is the discrete auto-correlation function of 
$\Tilde{\alpha}_a(m)$ defined by:
\begin{eqnarray}
    \Gamma^{\Tilde{\alpha}}_{a,a}[l]&=&\sum_{m=0}^{l_{\max}}\Tilde{\Tilde{\alpha}}_a[m]\Tilde{\Tilde{\alpha}}_a[m-l]^*\\
    &=&e^{-j2\pi\frac{a}{M}l}\times\sum_{m=0}^{l_{\max}}\Tilde{\Tilde{\alpha}}[m]\Tilde{\Tilde{\alpha}}[m-l]^*
\end{eqnarray}
with $\Tilde{\Tilde{\alpha}}_a[m]=\Tilde{\Tilde{\alpha}}[m]e^{-j2\pi\frac{a}{M}m}$ and:
\begin{equation}
\label{eq:alphatta}
\Tilde{\Tilde{\alpha}}[m] = \begin{cases}
 \Tilde{\alpha}(i) & \text{for } m = k_i, \quad i = 0,\ldots,K-1 \\
        0 & \text{otherwise.}
        \end{cases}
\end{equation}

Note that $\Gamma^{\Tilde{\alpha}}_{a,a}[l]$ is equal to zero for the lag  $l$ not in the set $\{k_m-k_n\}, \forall (m,n) \text{ with } m, n\in\{0,\ldots K-1\}$.  
\begin{SampleEnv}
Let's consider the following couples of channel parameters ($k_i,\,\tilde{\alpha}(i)$) equal to $(0,\,1.0)$, $(2,\,0.8)$ and $(3,\,0.5)$ for $i=0$, $1$ and $2$, respectively.
$\Gamma^{\Tilde{\alpha}}_{a,a}[l]$ can be evaluated in MATLAB/GNU Octave code for the example channel as:\\
{\small{\texttt{> alp\_tt=[1.0 0.0 0.8 0.5];}}}\\
{\small{\texttt{> alpa\_tt=alp\_tt.*exp(-2i*pi*a/M*(0:3));}}}\\
{\small{\texttt{> raa\_alp=conv(alpa\_tt,conj(fliplr(alpa\_tt));}}}\\[.2em]
In this example,  $\Gamma^{\Tilde{\alpha}}_{a,a}[l]$ is non-zero for $l=-3,\ldots,3$ because for each lag $l$, at least one product $\Tilde{\Tilde{\alpha}}[m]\Tilde{\Tilde{\alpha}}[m-l]^*$ is non-zero.\\
\label{Example1}
\end{SampleEnv}
 
By using (\ref{eq:Ca3}) in (\ref{eqMFFilter}) the DFT output of $\Tilde{z}_{a,a}[k]$ leads to: 
\begin{eqnarray}
\tilde{Z}_{a,a}[n]&=&M\Gamma^{\Tilde{\alpha}}_{a,a}[0]\delta[n-a] + M \sum_{l\neq0}\Gamma^{\Tilde{\alpha}}_{a,a}[l]\delta[n-a+l]\nonumber\\
&+&\tilde{W}_a[n]\hspace{1em}\text{ for }n=0,\ldots,M-1\label{eq:Zan}
\end{eqnarray}
where $\Gamma^{\Tilde{\alpha}}_{a,a}[0]$ corresponds to the channel energy: 
\begin{equation}
    \Gamma^{\Tilde{\alpha}}_{a,a}[0]=\sum_{m=0}^{K-1}|\alpha(m)|^2
\end{equation}
and the noise $\tilde{W}_a[n]$  at the DFT output is a complex circular Gaussian discrete stochastic process with the auto-correlation function given by:
$E[\tilde{W}_a[n]\tilde{W}_a^*[n-l]]=\sigma_w^2\Gamma^{\Tilde{\alpha}}_{a,a}[l]$.

Note that, thanks to the channel energy term in $n=a$, it allows us to consider naturally the coherent detection scheme via the real part of the DFT output in (\ref{eq:Zan}) for detecting the LoRa symbol which provides SER performance gain in comparison with the non-coherent detection because the noise variance for the coherent detection (real part) is half of the noise variance for the non-coherent detection (magnitude).

The symbol detection of the ideal-MF detector is then:
\begin{eqnarray}
  \widehat{a} = \underset{n}{\argmax} \quad \Re \{ \Tilde{Z}_{a,a}[n] \}.
  \label{eq:aest idela-mf}
\end{eqnarray}

From (\ref{eq:Zan}) we observe that the main peak $\Gamma^{\Tilde{\alpha}}_{a,a}[0]$  (peak of interest) is at index frequency $n=a$, and the parasitic peaks $\Gamma^{\Tilde{\alpha}}_{a,a}[l]$ are symmetric from $n=a$ at the distances $l=k_m-k_n$ for all $m\neq n$.

The ideal-MF approach aims to maximize the SNR at the symbol index frequency $n=a$ of the DFT output \eqref{eq:Zan}.
The SNR output of $\Re\{\Tilde{Z}_{a,a}[n]\}$  at the frequency index $n=a$ is then given by:
\begin{equation}
\label{eq:SNRa}
    \SNR_{a}=\frac{2M}{\sigma^2}\sum_{m=0}^{K-1}|\alpha(m)|^2.
\end{equation}

Hence, only by adding the MF step (i.e. multiplication by the channel coefficient $C_a^*[k]$ in (\ref{eqMFFilter0})) allows us to 1) use the coherent detection scheme and 2) improve the SNR output at $n=a$ from $\frac{2M}{\sigma^2}|\alpha(0)|^2$ (with the original system (\ref{eq:Rak2})) to (\ref{eq:SNRa}) by exploiting the channel energy.

\begin{figure}[tbp]
  \centering
  \vspace{-1em}
  \includegraphics[width=0.49\textwidth]{figures/plot_ratio_snr_mf_v2.eps}
  \caption{Comparison of $\Delta_a^{\coh}$ and $\Delta_a^{\nocoh}$ (original coherent and non-coherent systems, respectively) with $\Delta_a^{\idealmf}$ (ideal-MF) and $\Delta_a^{\mf}$ (MF), $\SF=7$.}
\label{fig:plot_tatio_snr}
\end{figure}

Performance of the ideal-MF is also strongly dependent on the biggest amplitude of the parasitic peaks. 
The higher parasitic peaks at the DFT output could be selected in presence of noise and lead to an erroneous detected symbol. 
By considering the performance indicator $\Delta_a$  as the ratio between the higher parasitic peak over the peak of interest (i.e. $n=a$), we obtain 
for
the original coherent and non-coherent systems 
in \eqref{eq:Rak2}: 
\begin{subequations}
\begin{align}
\Delta_a^{\coh} &= \underset{l \neq 0}{\max}  \quad \Re\{\Tilde{\alpha}_a(l)\}/{\alpha}(0) \\
\Delta_a^{\nocoh} &= \underset{l \neq 0}{\max}  \quad |\Tilde{\alpha}_a(l)|/{\alpha}(0)
\end{align}
\end{subequations}
 and for the ideal-MF system in \eqref{eq:Zan}:
\begin{equation}
\Delta_{a}^{\idealmf} = {\max_{l \neq 0}\Re\{\Gamma^{\Tilde{\alpha}}_{a,a}[l]\}}/{\Gamma^{\Tilde{\alpha}}_{a,a}[0]}.
\label{eq:Delta_a_mf}
\end{equation}

The smaller the ratio, the better the SER performance will be, because the peak of interest at $n=a$ will be more discriminated in presence of noise.
For the coherent cases (legends: ``original-coh.", ``ideal-MF" and ``MF" in Fig.~\ref{fig:plot_tatio_snr}), we observe $\Delta_a$ should also depend on the value of the transmitted symbol $a$, and so symbol error detection should also depends on $a$.
We considered the C1-channel in \eqref{eq:channel_example} as example. Note that the plot ``MF"  on the figure is about the MF receiver that will be detailed in the following subsec.~\ref{subsec:RakeMF}.

For the non-coherent detection scheme of the original system (legend ``original non-coh."), the  ratio $\Delta_a^{\nocoh}$ is constant and equals to 0.8.
It is the worst case because it is the largest value which is achieved whatever the transmitted symbol $a$.
For the original coherent detection scheme (legend ``original coh.") the ratio is depending on $a$ to reach the maximum at 0.8 for some particular values of $a$.
The resulting symbol error detection for a given transmitted symbol $a$ is $a$-dependent and we can clearly deduce better results for the original coh. over original non-coh. as SER performance is averaged over all symbols.
In the same way, we clearly observe that $\Delta_{a}^{\idealmf}$ (legend ``ideal-MF") is lower than $\Delta_{a}^{\coh}$ 
at the higher values that dominate the detection error of the symbol. The  performance indicator $\Delta_a$ shows that
better SER performance is expected with the ideal-ML.

The ratio between the maximum values of original-coh. and ideal-ML, which is indicated by the double arrow in Fig.~\ref{fig:plot_tatio_snr}, is equal to $\sum_m |\alpha(m)|^2=1.89$.
We conclude that the channel energy will drive the gain in terms of detection performance of the MF detector in comparison with the original coh. 
In other words, if additional channel paths of MPC do not carry significant energy, the interest of MF becomes very limited because the original coh. will bring close performance to the flat-fading AWGN channel.

\subsection{MF or RAKE receiver}\label{subsec:RakeMF}

We may see in Section~\ref{subsec:IdealMF} that the ideal-MF receiver depends on the transmitted symbol $a$ via $C_a^*[k]$ which is not available in practice. 
A way to circumvent this problem is to perform an exhaustive search with all the possible symbols $ b \in \{0,\ldots,M-1\}$ as follows:
\begin{equation}
        \Tilde{z}_{a,b}[k] = C_{b}^*[k] \Tilde{r}_a[k].
    \label{eqMFFilterCand}
\end{equation}

Note that subscript $b$ used in $\Tilde{z}_{a,b}[k]$ is to remind that the MF coefficient is $C_{b}^*[k]$, whereas the subscript $a$ is to remind that the current transmitted symbol is $a$, which is used in the received DC signal $\Tilde{r}_{a}[k]$.

The estimated symbol is the candidate that maximizes the DFT output of $\Tilde{z}_{a,b}[k]$  at the index frequency $b$:
\begin{eqnarray}
  \widehat{a} = \underset{b}{\argmax} \quad \Re \{ \Tilde{Z}_{a,b}[b] \}
  \label{eq:argmaxZb}
\end{eqnarray}
where:  
\begin{equation}
\Tilde{Z}_{a,b}[b]=\sum_{k=0}^{M-1} \Tilde{z}_{a,b}[k]e^{-j2\pi bk/M}.
\label{eq:TFD Zbi}
\end{equation}

The detector \eqref{eq:argmaxZb} is only denoted MF because the knowledge of $C_a[k]$ is not necessary as for the ideal-MF.
By following the same development used for the ideal-MF in Section~\ref{subsec:IdealMF}, the DFT output of the MF is: 
\begin{eqnarray}
\tilde{Z}_{a,b}[b]&=&M\Gamma^{\Tilde{\alpha}}_{a,b}[0]\delta[b-a] + \sum_{l\neq0}M\Gamma^{\Tilde{\alpha}}_{a,b}[l]\delta[b-a+l]\nonumber\\
&&+\tilde{W}_b[b]\label{eq:Zbn}
\end{eqnarray}
where $\Gamma^{\Tilde{\alpha}}_{a,b}[l]$ is the cross-correlation function of $\Tilde{\alpha}_a(m)$ and $\Tilde{\alpha}_b(m)$ defined by:
\begin{eqnarray}
\label{eq:cross_corr}
    \Gamma^{\Tilde{\alpha}}_{a,b}[l]&=&\sum_{m}^{}\Tilde{\Tilde{\alpha}}_a[m]\Tilde{\Tilde{\alpha}}_b[m-l]^*.
\end{eqnarray}

\begin{figure}[ht]
  \centering
  \includegraphics[width=0.49\textwidth]{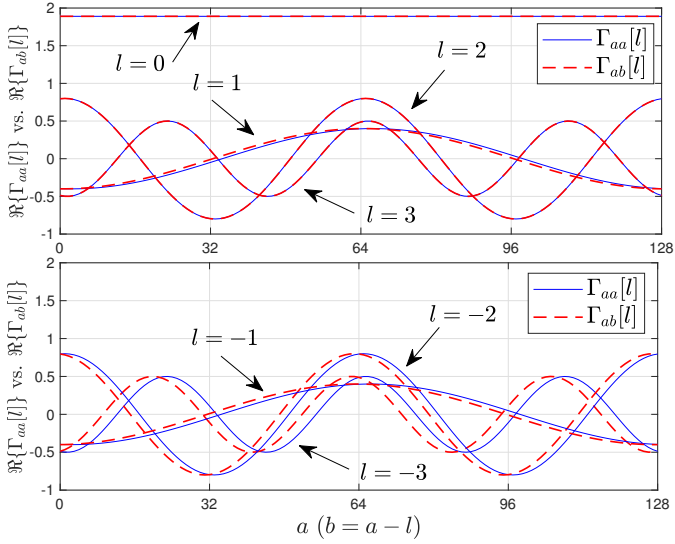}
  \vspace{-2em}
  \caption{Comparison of peak amplitudes between ideal-MF  (i.e. auto-correlation $\Re\{\Gamma^{\tilde{\alpha}}_{a,a}[l]\}$) and MF (i.e. cross-correlation $\Re\{\Gamma^{\tilde{\alpha}}_{a,b}[l]\}$) 
  for C1-channel as a function of
  symbol $a$ (x-axis).  $\Re\{\Gamma^{\tilde{\alpha}}_{a,b}[l]\}$ is evaluated at the peak locations $b=(a-l)$ mod $M$ for $l=0,\pm1,\pm2,\pm3$.}
\label{fig:Zb_a}
\end{figure}

The MF receiver is not equivalent to the ideal-MF receiver (see (\ref{eq:aest idela-mf}) and (\ref{eq:argmaxZb}) for comparison).
The difference between $\Tilde{Z}_{a,a}[b]$ for the ideal-MF and $\Tilde{Z}_{a,b}[b]$ for the MF  appears only for $\Gamma^{\Tilde{\alpha}}_{a,a}[l]$ in (\ref{eq:Zan}) which is replaced by 
$\Gamma^{\Tilde{\alpha}}_{a,b}[l]$ in (\ref{eq:Zbn}). 
The main peak at $b=a$  is the same but amplitude of parasitic peaks at $b=a-l$ ($l\neq0$) are different ($\Gamma^{\Tilde{\alpha}}_{a,a}[l]$ vs. $\Gamma^{\Tilde{\alpha}}_{a,b}[l]$).
Otherwise, the noise statistic of $\Tilde{W}_a[b]$ or $\Tilde{W}_b[b]$ is equivalent with the same power 
$\sigma_w^2\sum_m|\alpha(m)|^2$. Even if the two tests  $\Tilde{Z}_{a,a}[b]$ and  $\Tilde{Z}_{a,b}[b]$ are not identical, SER performance will be very close
because the difference between  $\Gamma^{\Tilde{\alpha}}_{a,a}[l]$  and 
$\Gamma^{\Tilde{\alpha}}_{a,b}[l]$ involves only a shifting of the peak values as shown in Fig.~\ref{fig:Zb_a}.  
We considered the C1-channel in \eqref{eq:channel_example}.
Fig.~\ref{fig:Zb_a} compares $\Gamma^{\Tilde{\alpha}}_{a,a}[l]$ with $\Gamma^{\Tilde{\alpha}}_{a,b}[l]$ for each symbol $a$. $\Gamma^{\Tilde{\alpha}}_{a,b}[l]$ needs to be evaluated at the peak locations $b=a-l$ mod $M$ for $l=0,\,\pm1,\,\pm2,\,\pm3$. 
We observe for $l=0$, 2 and 3,  $\Gamma^{\Tilde{\alpha}}_{a,a}[l]=\Gamma^{\Tilde{\alpha}}_{a,b}[l]$ and for $l=-3$, $-2$, $-1$ and 1, $\Gamma^{\Tilde{\alpha}}_{a,b}[l]$  is a shifted version of $\Gamma^{\Tilde{\alpha}}_{a,a}[l]$.
By considering the average performance over all the possible transmitted symbols $a$, SER performance will be very close because peak values are the same (only shifted).
We conclude that ideal-MF (known symbol $a$) and MF (unknown $a$) approaches lead to similar performances. 
The performance indicator for the MF receiver is given by:
\begin{equation}
\Delta_{a}^{\mf} = {\underset{\substack{l \neq 0\\\text{and:}~b=a-l\text{ mod }M}}{\max} \Re\{\Gamma^{\Tilde{\alpha}}_{a,b}[l]\}}/{\Gamma^{\Tilde{\alpha}}_{a,b=a}[0]}.
\label{eq:Delta_a_mf_true}
\end{equation}

In Fig.~\ref{fig:plot_tatio_snr} we reported also $\Delta_{a}^{\mf}$ to compare with  $\Delta_{a}^{\idealmf}$. We clearly observe very slight differences that will produce similar performance in term of SER, which is confirmed by simulations.


\begin{figure}[ht]
  \centering
  \input{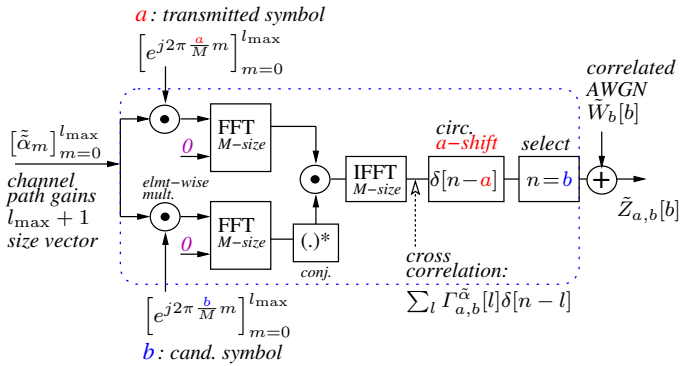}
  \caption{Equivalent system model \eqref{eq:Zbn} for the MF receiver and by considering the received signal approximation given in (\ref{eq:ykDech2}).}
\label{fig:equiv}
\end{figure}

From~\eqref{eq:Zbn}  an equivalent system model for MF receiver can be summarized in Fig.~\ref{fig:equiv}.
It corresponds to computing the  cross-correlation function \eqref{eq:cross_corr} via $M$-size FFT/IFFT, the $M$-size IFFT output is then (right) circular shifted by $a$, and the index $n=b$ (candidate symbol) is selected.
It is interesting to note that this equivalent system model can be used to perform simulations in a very fast way. The processing in the dash box on Fig.~\ref{fig:equiv} can be done off-line where the noise-free values of $\Tilde{Z}_{a,b}[b]$ can be stored in a matrix of size $M\times M$ for each transmitted symbol $a$ (row index) and for each candidate symbol $b$ (column index).
In on-line simulations, for each transmitted symbol $a$ we just have to select the corresponding row into the stored matrix and to add the correlated noise realization $\Tilde{W}_b[b]$ for $b=0,\ldots,M-1$ to obtain \eqref{eq:Zbn}.

By using (\ref{eqMFFilterCand}) and (\ref{eq:path}) (with $a$ replaced by $b$ in~\eqref{eq:path}) into (\ref{eq:TFD Zbi}), the statistic $\Tilde{Z}_{a,b}[b]$ can be also expressed like a RAKE receiver structure as follows:
\begin{eqnarray}
\label{eq:Zb_b}
\Tilde{Z}_{a,b}[b]&=&\sum_{i=0}^{K-1}\Tilde{\alpha}_b^*(i) \sum_{k=0}^{M-1} \Tilde{r}_a[k] e^{-j2\pi\frac{k}{M}(b-k_i)}\\
&=&\sum_{i=0}^{K-1}\Tilde{\alpha}_b^*(i) \Tilde{R}_a[b-k_i].
\label{eq:Zb_b2}
\end{eqnarray}

Note that for $b=a$, by using (\ref{eq:Rak2}) in (\ref{eq:Zb_b2}) we retrieve the channel energy:
$\Tilde{Z}_{a,a}[a]=\sum_{i=0}^{K-1}|\Tilde{\alpha}_a(i)|^2+W_a[a]=\sum_{i=0}^{K-1}|\alpha(i)|^2+W_a[a]$, that is the principle of the RAKE receiver.
The MF approach can be seen also as a RAKE receiver.

The ideal-MF can be also retrieved from the RAKE receiver structure as follows:
\begin{eqnarray}
\label{eq:Za_b}
\Tilde{Z}_{a,a}[b]
&=&\sum_{i=0}^{K-1}\Tilde{\alpha}_a^*(i) \Tilde{R}_a[b-k_i].
\label{eq:Za_b2}
\end{eqnarray}

\subsection{Candidate MF (cand-MF) or candidate RAKE (cand-RAKE) receiver}\label{subsec:candMFRake}

The main problem of the MF or RAKE receiver is the computational complexity due to the exhaustive search in the set of possible symbols of $M=2^{\SF}$ elements.
To address this issue, we propose to estimate $a$, chosen from a list of most probable candidate symbols.
The set of candidate symbols can be created with two different approaches. \\
\subsubsection{Fixed candidate number}\label{subsubsec:fixed_cand_number}
The first approach consists in selecting the $N_c$ frequency bins having the highest magnitude in $|\Tilde{R}_a[n]|$, forming the candidate set $\widehat{a} \in \mathcal{A} = \{b_0,b_1,\ldots,b_{N_c-1}\}$.
Indeed, from (\ref{eq:Rak2}) the highest peak magnitudes $|\alpha(i)|$ are at the location bins $n = a-k_i$, the biggest corresponding to the correct symbol $a$ ($k_0=0$).
In presence of AWGN noise, a sufficient large set of candidates allows to catch the symbol of interest.
Moreover, this method enables the control of the number of candidates to use driven by $N_c$ parameter and is more suited to evaluate $N_c$ impact on SER performance.
However,  this approach significantly increases the complexity of the  receiver as it requires to use a sorting algorithm.\\

\begin{table}[ht]
 \setlength{\tabcolsep}{2pt}
    \centering
\begin{tabular}{c||c|c}
\begin{tabular}{l}MF vs.\\cand-MF\end{tabular} & cmult     &  cadd \\\hline\hline
$\Tilde{\alpha}_b(i)$ (\ref{eq:alphat_a}) & $KM$ vs. $KN_c$ & - \\\hline
$C_b[k]$ (\ref{eq:path}) & $KM^2$ vs. $KMN_c$ & $(K-1)M^2$ vs.$(K-1)MN_c$  \\\hline
$\tilde{z}_{a,b}[k]$ (\ref{eqMFFilterCand})  & $M^2$ vs.  $MN_c$ & - \\\hline
$\Tilde{Z}_{a,b}[b]$ (\ref{eq:TFD Zbi}) & $M^2$ vs. $MN_c$ & $(M-1)M$ vs. $(M-1)N_c$\\\hline
 \begin{tabular}{c}tot.~MF\\tot.~cand-MF\end{tabular} & \begin{tabular}{c}$M(2M+KM+K)$\\$N_c(2M+KM+K)$\end{tabular} &   \begin{tabular}{l}
 $M(MK-1)$\\$N_c(MK-1)$
\end{tabular} \\[1em]
\end{tabular}
  \caption{MF and candidate MF (cand-MF) complexity in terms of number of complex multiplications (cmult) and complex additions (cadd).}
    \label{tab:mf}
\end{table}

\begin{table}[ht]
 \setlength{\tabcolsep}{2pt}
    \centering
\begin{tabular}{c||c|c}
\begin{tabular}{l}RAKE vs. \\cand-RAKE\end{tabular}  & cmult     &  cadd \\\hline\hline
\begin{tabular}{c}$\Tilde{R}_a[n]$ \\(e.g. FFT radix-2)\end{tabular}  & $\frac{M}{2}\log_2M$  & $M\log_2M$ \\\hline
$\Tilde{\alpha}_b(i)$ (\ref{eq:alphat_a}) & $KM$ vs. $KN_c$ & - \\\hline
$\Tilde{Z}_{a,b}[b]$ (\ref{eq:Zb_b2}) & $KM$ vs. $KN_c$ & $(K-1)M$ vs. $(K-1)N_c$\\\hline
 \begin{tabular}{c}tot.~RAKE\\tot.~cand-RAKE\end{tabular} & \begin{tabular}{c}$\frac{M}{2}\log_2 M+2KM$\\$\frac{M}{2}\log_2 M+2KN_c$\end{tabular} &   \begin{tabular}{l}$M\log_2 M+(K-1)M$\\
$M\log_2 M+(K-1)N_c$\end{tabular} \\[1em]
\end{tabular}
 \caption{RAKE and Candidate RAKE (cand-RAKE)  complexity in terms of number of complex multiplications (cmult) and complex additions (cadd).}
    \label{tab:rake}
\end{table}

\subsubsection{Variable candidate number}\label{subsubsec:variable_cand_number}
The second approach addresses the complexity issue of method 1 by using a threshold to select the candidates.
This threshold can be designed as a fraction of the maximum value in $|\Tilde{R}_a[n]|$:
\begin{equation}
    \begin{split}
        \lambda_c = \rho_c \times \max_n \quad \left|\Tilde{R}_a[n]\right|
    \end{split}
    \label{eqThresholdCand}
\end{equation}
where $\rho_c \in \ [0,1[$ is the arbitrary fractional magnitude.
The candidates are then the DFT magnitude bins that are above~$\lambda_c$:
\begin{equation}
    \mathcal{A} = \left\{ n~: \left|{\Tilde{R}_a}[n]\right|> \lambda_c \right\}.
    \label{eqCandSelectThres}
\end{equation}

The value of $\rho_c$ drives the SER and computation complexity trade-off.
The lower $\rho_c$, the larger is the set $\mathcal{A}$.
This allows to catch the symbol of interest with high probability, which improves SER performance but at the expense of higher complexity.
A high $\rho_c$ value will have the opposite effect.\\

Once the set of candidate symbols is obtained, the symbol detection is:
\begin{eqnarray}
 \widehat{a} = \underset{b_u \in \mathcal{A}}{\argmax} \quad \Re \{ \Tilde{Z}_{a,b_u}[b_u] \},
  \label{eq:argmaxZbi}
\end{eqnarray}
where the test $\Tilde{Z}_{a,b_u}[b_u]$ can be evaluated via the MF approach by using \eqref{eqMFFilterCand} and \eqref{eq:TFD Zbi} or via the RAKE approach in \eqref{eq:Zb_b2}. 

\begin{figure*}[tb]
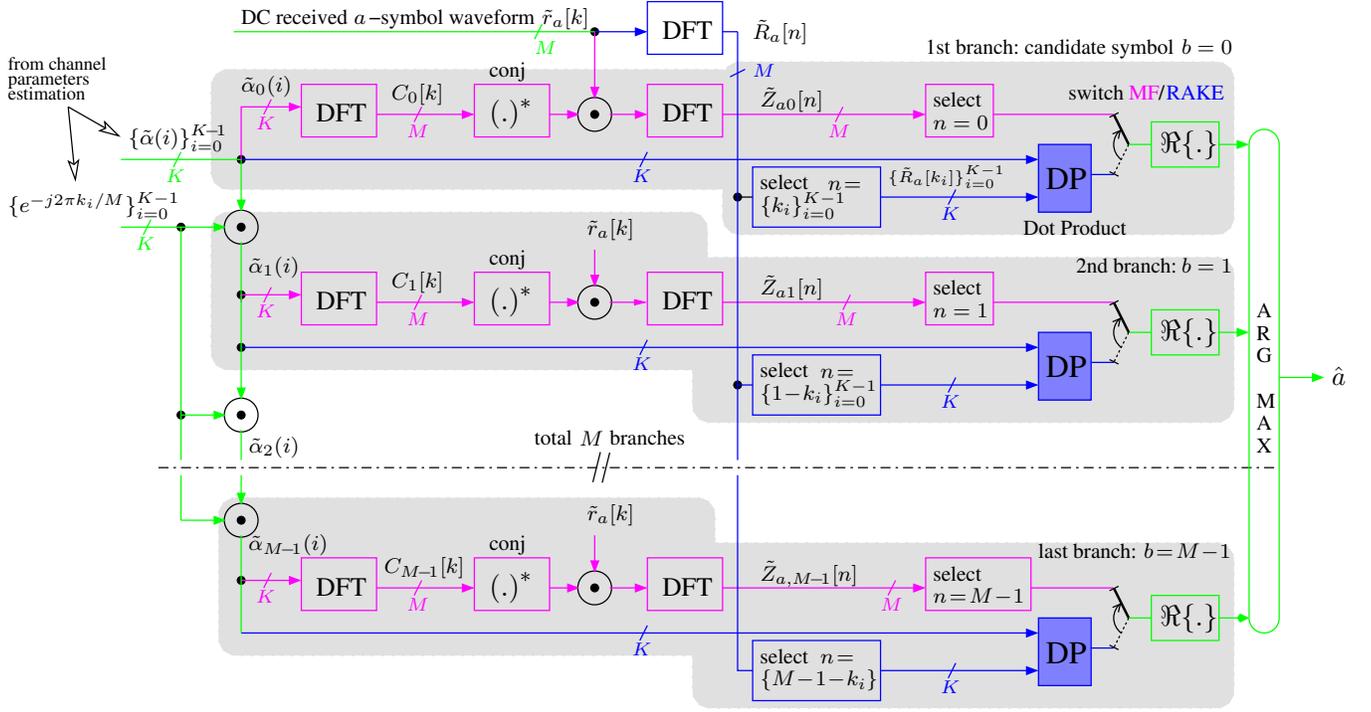

  \centering
  \input fig3.pspdftex
  \caption{Block diagram of the MF (magenta) and RAKE (blue) detectors. The
    green color is used for both detectors. The cand-MF or cand-RAKE consists in
    computing only the branches corresponding to ${\cal A}$.
    }
\label{fig:RAKE MF}
\end{figure*}

Fig. \ref{fig:RAKE MF} presents a visual comparison between MF and RAKE.
The MF specific operations are denoted with magenta color while RAKE ones are in blue color.
In the figure, we suppose that the channel parameters estimation $\Tilde{\alpha}(i)$, $k_i$ and $K$ have been already derived (see Section \ref{sec:ChannelEst}).
The branches for each candidate symbol $b=0, 1,\ldots, M-1$ are highlighted with gray boxes.
We may also see that the RAKE receiver performs exclusively in the frequency domain, the latter is indeed a modified version of the LoRa legacy coherent receiver, while MF executes the first part of its front-end in the time domain (\emph{i.e.}, $C_b^*[k]\tilde{r}_a[k]$), see the operations on the left of each DFT boxes in the figure.

\subsection{Complexity evaluation}
We assess in the section the computational complexity of MF and RAKE receivers in terms of complex additions and multiplications (denoted cadd and cmult, respectively) and execution time. \\
\subsubsection{Complexity evaluation in terms of complex operations}\label{subsubsec:complexity_op}

Tab.~\ref{tab:mf} and Tab.~\ref{tab:rake} show the number of cadd and cmult required for MF and RAKE receivers, respectively.
We may see that RAKE approach presents less computational complexity than MF.
The gain in complexity of RAKE comes from the fact that $C_{b}[k]$ and $\tilde{z}_{a,b}[k]$ are not necessary to be evaluated and an FFT algorithm can be used to compute $\Tilde{R}_a[n]$ in 
\eqref{eq:Zb_b2}.

\begin{figure}[ht]
  \centering
    \includegraphics[width=0.49\textwidth]{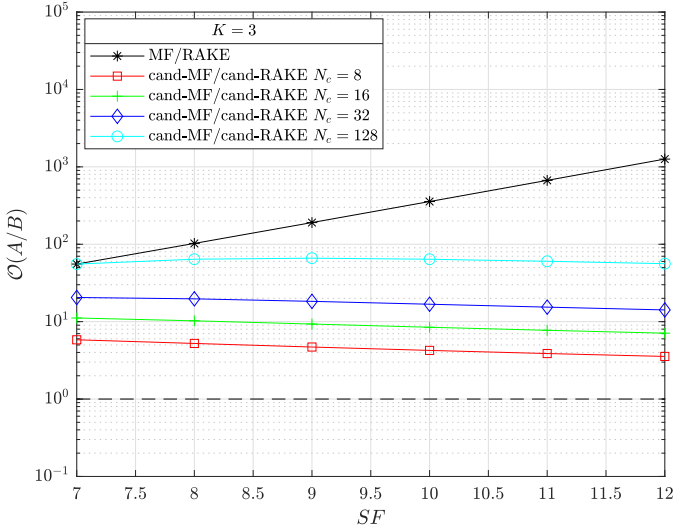}
  \caption{Complexity ratios $\mathcal{O}$(MF/RAKE) and $\mathcal{O}$(cand-MF/cand-RAKE) as a function of SF and $N_c$, $K=3$.}
  \label{fig:MF_RAKE_CAND_MF_CAND_RAKE_complexity_eq}
\end{figure}

Fig.~\ref{fig:MF_RAKE_CAND_MF_CAND_RAKE_complexity_eq} presents graphically the complexity ratios  $\mathcal{O}$(MF/RAKE) and 
$\mathcal{O}$(cand-MF/cand-RAKE) as a function of SF, $N_c$ and for $K=3$.
The complexity ratio $\mathcal{O}({A/B})$ is defined by the sum of total number of cadd and cmult of $A$ receiver over the sum of total number of cadd and cmult of $B$ receiver:
\begin{equation}
    \mathcal{O}{(A/B)} = \frac{\left(\text{tot.~cadd} + \text{tot.~cmult}\right)_A}{\left(\text{tot.~cadd} + \text{tot.~cmult}\right)_B}.
    \label{eq:tot_cadd_cmult_A_B}
\end{equation}

We consider that a complex multiplication has the same complexity as a complex addition, thanks to dedicated instructions set in hardware.
The following results can be drawn from Fig.~\ref{fig:MF_RAKE_CAND_MF_CAND_RAKE_complexity_eq}.
First, RAKE receiver outperforms MF in terms of complexity, with a log-ratio that grows linearly with SF.
This represents a huge complexity gain, especially for high SF values ($> 10^3$ at $\SF = 12$).
And secondly, by focusing on ``candidate" approach (i.e. cand-MF vs. cand-RAKE), the behavior looks different because
the determining parameter is $N_c$ whereas the complexity gain is almost constant over SF.
We conclude that the RAKE implementation requires much less computational complexity in terms of complex operations (cadd and cmult) than the MF implementation. \\

\begin{algorithm}[ht]
\DontPrintSemicolon
\SetKwInOut{Input}{inputs}
\Input{
$\tilde{\alpha}(i)$ and $k_i$ for $i=0,\ldots,K-1$ (channel parameters assumed known), 
$\mathbf{r}=\{\Tilde{r}_{a}[k]\}_{k=0}^{M-1}$ the received DC signal vector, and $\rho_c$ the threshold for candidates selection.}
${\mathbf{R}}:=\operatorname{FFT}({\mathbf{r}})$ \;
$R_{\max}:=\max(\operatorname{abs}(\mathbf{R}))$\;
${\mathcal A}:=\operatorname{find}(\operatorname{abs}(\mathbf{R})>\rho_cR_{\max}$) \hfill \textcolor{teal}{\%find indexes}\;
$\mathbf{c}:=\mathbf{0}_M$\hfill \textcolor{teal}{\%init $M$-size vector}\;
 \ForEach{candidate $b_u\in \mathcal A $}{
 \For{$i=0$ to $K-1$}{
$\mathbf{c}\left[k_i\right]:=\Tilde{\alpha}(i)\exp(-j2\pi k_i b_u/M)$\;
}
${\mathbf{C}}:=\operatorname{FFT}({\mathbf{c}})$ \hfill \textcolor{teal}{\%$C_{b_u}[k]$ in (\ref{eq:path})} \;
\For{$k=0$ to $M-1$}{
$\mathbf{z}[k] := \mathbf{r}[k]\mathbf{C}^*[k]$ \hfill \textcolor{teal}{\%$\Tilde{z}_{a,b_u}[k] = C_{b_u}^*[k] \Tilde{r}_{a}[k]$}\;
}
${\mathbf{Z}}:=\operatorname{FFT}({\mathbf{z}})$ \;
store $v_u := \Re\{\mathbf{Z}[b_u]\}$\hfill \textcolor{teal}{\%$\Tilde{Z}_{a,b_u}[b_u]$ in \eqref{eq:TFD Zbi}}\;
}
\textbf{return}  $\widehat{a} = \underset{u \rightarrow b_u}{\argmax} \quad v_u$\;
\caption{cand-MF symbol estimation}
 \label{algo:candMF}
\end{algorithm}

\begin{algorithm}[ht]
\DontPrintSemicolon
\SetKwInOut{Input}{inputs}
\Input{same as Algorithm 1}
${\mathbf{R}}:=\operatorname{FFT}({\mathbf{r}})$ \;
Determine $\mathcal A$ (same as Alg.~1, lines 2-3)\;
 \ForEach{candidate $b_u\in\mathcal A$}{
 $s:=0$\;
 \For{$i=0$ to $K-1$}{
$m := \mod(b_u- k_i,M)$\ \textcolor{teal}{\%modulo operation}\;
$y := \mathbf{R}[m]$;\;
$x:=\Tilde{\alpha}(i)\exp(-j2\pi k_i b_u/M)$\;
$s := s + yx^*$ 
\;
}
 store $v_u := \Re\{s\}$ \hfill \textcolor{teal}{\%$s=\Tilde{Z}_{a,b_u}[b_u]$ in \eqref{eq:Zb_b2}}\;
 }
 \textbf{return}  $\widehat{a} = \underset{u \rightarrow b_u}{\argmax} \quad v_u$\;
 \caption{cand-RAKE symbol estimation}
 \label{algo:candRAKE}
\end{algorithm}

\subsubsection{Complexity in terms of execution time}\label{subsubsec:complexity_exec}

Algorithms with candidate approach for both MF and RAKE are provided in Algorithms \ref{algo:candMF} and \ref{algo:candRAKE}, respectively.
The \textit{full} RAKE and MF correspond also to Algorithms \ref{algo:candMF} and \ref{algo:candRAKE}
with the set ${\cal{A}}=\{0,1,\ldots,M-1\}$ and the corresponding code in lines~1-3 (Alg.~1) and line~2 (Alg.~2) must be omitted.

\textcolor{black}{
Unlike the complexity study in terms of complex operations (cadd and cmult), complexity in terms of execution time takes into account the determination of {$\cal A$} and the $\argmax$ operation (Alg.~1, line 13 or Alg.~2, line 11)  in order to perform fair comparisons between MF vs. RAKE and, more particularly, between RAKE vs. cand-RAKE. Indeed, in the latter case, cand-RAKE adds computational burden to determine {$\cal A$} in comparison with RAKE where ${\cal A}=\{0,\ldots M-1\}$, and so the execution time is globally measured on the entire algorithms.
}

Note that, for MF, \eqref{eq:path} and \eqref{eq:TFD Zbi} can be actually implemented with FFT algorithm (Alg.~1, lines~8 and 11) to enable more computation efficiency.
We may see from Alg.~1 and Alg.~2 that cand-RAKE (Alg.~\ref{algo:candRAKE}) has a less complex demodulation scheme than cand-MF (Alg.~\ref{algo:candMF}).
\textcolor{black}{In fact, the loop over $M$ elements in Alg.~1 line 9 (and absent in Alg.~2) considerably slows down the execution.
Moreover, cand-MF requires two $M$-size FFT algorithm calls (lines 8 and 11) that further put a burden on complexity.}

\begin{figure}[ht]
    \centering
    \includegraphics[width=0.49\textwidth]{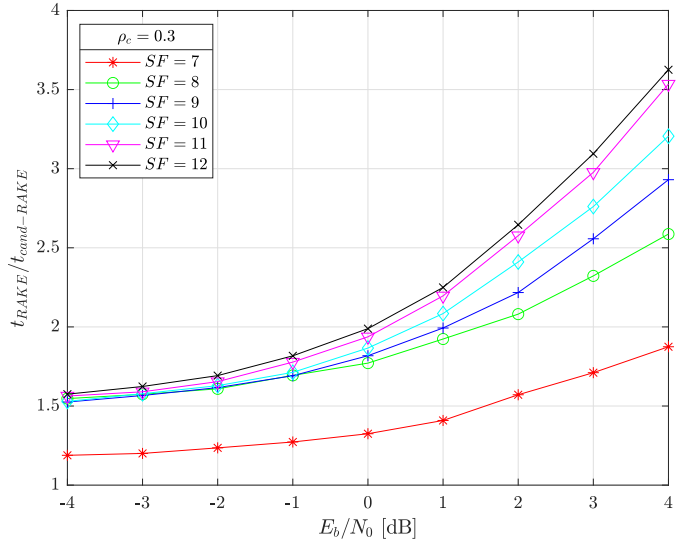}
    \caption{Execution time  ratio of RAKE over cand-RAKE compiled C-code (generated by MATLAB) as a function of $E_b/N_0$ for each SF.
 The variable number of candidate selection method is used with $\rho_c = 0.3$ that leads to the same cand-RAKE SER performance as \textit{full} RAKE.}
    \label{fig:RAKE_vs_candRAKE_ratio_mex}
\end{figure}

Fig.~\ref{fig:RAKE_vs_candRAKE_ratio_mex} presents execution time  comparison between RAKE and cand-RAKE as the execution time  ratio of C compiled RAKE and cand-RAKE algorithms (Alg.~2), as a function of SNR per bit, $E_b/N_0$ defined in \eqref{eq:SNREbNo}, and for each SF.
\textcolor{black}{The execution time is averaged over numerous trials to derive an average execution time.
Indeed, the number of selected candidates  (line~3, Alg.~1) fluctuates over different trials, due to AWGN realization.
That is, the loops in Alg.~1 line 5 and Alg.~2 line 3 are iterated over a different number of elements and thus impact the execution time.}
We choose 
$\rho_c = 0.3$ that gives almost same SER performance between RAKE and cand-RAKE
(see Figs.~\ref{fig:Simu_MF_RAKE_CAND_C1_SF7} and \ref{fig:Simu_MF_RAKE_CAND_C1_SF10} in the simulation part).
MF and cand-MF algorithms are slower so they are not presented in the figure.
We may see that increasing $E_b/N_0$ improves the execution time  gain of cand-RAKE as it actually reduces the size of $\mathcal{A}$ candidates set.
This is also true when increasing SF with a maximum gain of about~$3.6$.
\textcolor{black}{In conclusion, we clearly observe a gap between $\SF=7$ and $\SF>7$ that shows significant execution time gain for cand-RAKE over RAKE
at $\SF>7$ and, in particular at higher SNR values.}\\


We have seen that both MF and RAKE detectors 
are equivalent (i.e. Eqs.~\eqref{eq:TFD Zbi} and \eqref{eq:Zb_b2} are equivalent)
but complexity is in favor of RAKE.
From now on, we consider RAKE detector for the rest of the paper.

\section{Channel parameters estimation}\label{sec:ChannelEst}

For a proper demodulation, the receiver must estimate channel parameters: the tap delays $k_i$ and their associated path gains $\Tilde{\alpha}(i)$. Note that $\Tilde{\alpha}(i)$ is not equal to the channel path gain $\alpha(i)$ as shown in (\ref{eq:alpha_i_tilde}), but only the terms $\Tilde{\alpha}(i)$ are needed to compute the MF or RAKE tests. 
We assume that the receiver is synchronized on the arrival time of the first path (i.e. $k_0 = 0$).
To estimate $k_i$ and $\Tilde{\alpha}(i)$, we use pilot symbols located at the beginning of each frame and known by both the transmitter and the receiver.
Using pilot symbols implies that the channel coherence time is at least equal to the frame duration i.e. $T_{coh} \ge  N_f \times T$ with $N_f$ the total number of symbols in a frame.
In the literature, $a = 0$ is considered for pilot symbols to simplify equations.
The detailed LoRa frame structure can be found in \cite{bernier} for example but we use here without loss of generality the simplest frame structure: $N_p$ pilot symbols ($a=0$) followed by data symbols, i.e $N_f = N_p + N_d$.
$N_p=8$ is a typical value used in LoRa.

From (\ref{eq:Rak2}) with $a=0$, we propose to use the averaged DFT over the $N_p$ pilot symbols to improve path gains estimation of $\Tilde{\alpha}_0(i)=\Tilde{\alpha}(i)$.
The averaged DFT over $N_p$ realizations is given by:
\begin{equation}
    \begin{split}
        \left\langle{\Tilde{R}_0}[n]\right\rangle = \frac{1}{N_p} \sum_{p=0}^{N_p-1} \Tilde{R}_0^p[n]\hspace{1em}\text{for }n=0,\ldots,M-1
    \end{split}
    \label{eqPilotSymbolAvg}
\end{equation}
where $\Tilde{R}_0^p[n]$ is the $p$-th realization of $\Tilde{R}_0[n]$ (see \eqref{eq:Rak2} with $a=0$).
This way, the noise power is reduced by $N_p$ factor.
As we assume to be synchronized on the first path, i.e. $\widehat{k}_0 = 0$, and 
echoes arriving in interval lower than $k_{\max}\ge1$ sample periods, 
the receiver then detects the DFT bins  having magnitudes above a certain threshold, in the  range $n_{k_{\max}} = \{(M-k_{\max}),\ldots,M-1\}$:
\begin{eqnarray}
        n^{\prime} &=& \left\{ n_{k_{\max}}~: \left| \left\langle{\Tilde{R}_0}[n_{k_{\max}}] \right\rangle \right|> \lambda_p \right\}
    \label{eqTauEst2}
\end{eqnarray}
with:
\begin{equation}
    \begin{split}
\lambda_p = \rho_p \times  \left| \langle{\Tilde{R}_0}[0]\rangle \right|
    \end{split}
    \label{eqTThreshold}
\end{equation}
where $\rho_p \in \ ]0,1[$ is the arbitrary fractional magnitude.
The estimated path gains  $\widehat{\Tilde{\alpha}(i)}$ are then:
\begin{equation}
      \widehat{\Tilde{\alpha}(i)}=\left\langle{\Tilde{R}_0}[n^{\prime}]\right\rangle,
\end{equation}
with the relation between the raw DFT index $n^{\prime}$ and the estimated path delay $\widehat{k}_i$ given by:
\begin{equation}
   n^{\prime} = M-\widehat{k}_{i}.
\end{equation}

Once the couples $(k_i,\Tilde{\alpha}(i))$ are estimated, $\Tilde{\alpha}_{b_u}(i)$ can be computed with $\Tilde{\alpha}_{b_u}(i)=\Tilde{\alpha}(i) e^{-2j\pi k_i \frac{b_u}{M}}$.

\section{Comparison of RAKE and TDEL \cite{guo} detectors}

\subsection{TDEL detector overview} \label{subsec:TDEL}

The authors in \cite{guo} derived a simple detection scheme based on cyclic frequency correlation between the averaged preamble and each data symbols.
We denote this enhanced receiver as Time Delay Estimation LoRa (TDEL).
The main steps of TDEL are briefly described in what follows.
First, it averages the DFT over pilot symbols identically to our scheme for channel estimation.
It computes next the cyclic cross-correlation between averaged pilot and symbol DFT's as:
\begin{equation}
    \Gamma_{\Tilde{R}_0^{\prime},\Tilde{R}_a}[d] = \sum_{n=0}^{M-1} \left| \langle\Tilde{R}_0^{\prime}[n]\rangle \right| \left| \Tilde{R}_a[n+d\mod M] \right|
    \label{eq:CrossCorrTDEL}
\end{equation}
with $\Tilde{R}_0^{\prime}[n]$ having $\Tilde{R}_0[n]$ outputs ignored if below a certain threshold $\lambda_{\TDEL}$, i.e. $\Tilde{R_0^{\prime}}[n] = 0$ if $|\Tilde{R}_0[n]| < \lambda_{\TDEL}$.
$\lambda_{\TDEL}$ is defined as:
\begin{equation}
    \lambda_{\TDEL} = \rho_{\TDEL} \times \max_n \left| \Tilde{R}_0[n] \right|.
\end{equation}

Finally, TDEL chooses the frequency index that maximizes $\Gamma_{\Tilde{R}_0^{\prime},\Tilde{R}_a}[d]$ in \eqref{eq:CrossCorrTDEL} as the estimated symbol:
\begin{equation}
    \widehat{a} = \argmax_d\Gamma_{\Tilde{R}_0^{\prime},\Tilde{R}_a}[d].
\end{equation}

\subsection{RAKE and TDEL comparison} \label{subsec:RAKETDELCompar}

\begin{figure*}[tb]
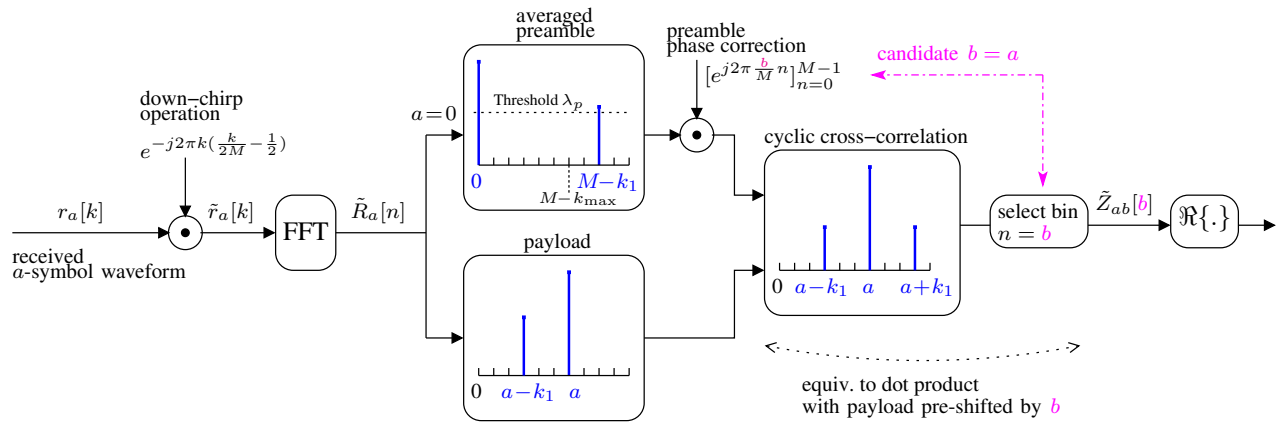

  \centering
  \input fig1.pspdftex
  \caption{Block diagram of the RAKE statistic $\tilde{Z}_{ab}[b]$ illustrated
    for two paths channel (with path delay $k_1$) and for the correct candidate
    symbol~$b=a$.
    This diagram is presented like a coherent version of the TDEL (see Fig.1 in \cite{guo}).
    The RAKE (resp. cand-RAKE) detector consist in computing $\tilde{Z}_{ab}[b]$ for each $b$ in $\{0,\ldots,M-1\}$ (resp.  $b\in{\cal A}$).
    The $\argmax$ operator is then computed on $\Re\{\tilde{Z}_{ab}[b]\}$ over all the candidate symbols~$b$ to detect the transmitted $a$-symbol.}
\label{fig:RAKEcoh}
\end{figure*}

In Fig.~\ref{fig:RAKEcoh} the RAKE receiver is presented in a similar way to the TDEL receiver (Fig.~1 in \cite{guo}) in order to highlight the similarities and differences of these two schemes.  

First, preamble waveform ($N_p$ symbols with $a=0$) is received.
The channel parameter estimation method presented in Section~\ref{sec:ChannelEst} is equivalent to average $\tilde{R}_0[n]$ and select the significant peaks over the threshold $\lambda_p$ (all the other frequency bins are forced to zero as it is illustrated in the figure). 
Once the couples $(k_i,\tilde{\alpha}(i))$ of the channel parameters are estimated, the preamble phase correction is performed thanks to the candidate $b$-symbol to obtain $\tilde{\alpha}_b(i)$.
Note that element-wise multiplication of the averaged preamble by $[e^{j2\pi n b/M}]_{n=0}^{M-1}$ in the figure is equal to $\tilde{\alpha}_b(i)=\tilde{\alpha}(i)e^{-j2\pi k_i b/M}$, which is required for the RAKE.

Second, the current $a$-symbol waveform is received in the payload.
$\tilde{R}_a[n]$ exhibit peaks of interest with phases depending on the current symbol $a$, and
noise peaks (not illustrated in the figure). Note that if $b=a$ (candidate symbol is correct), phases of  $\tilde{\alpha}_b(i)$ match with those in 
$\tilde{R}_a[n]$ at the corresponding frequency bins. Here comes the main difference with TDEL because in the TDEL scheme the magnitude of $\tilde{R}_a[n]$ is first performed (see \eqref{eq:CrossCorrTDEL} or Fig.1 in \cite{guo}), which leads to a non-coherent receiver where the phase information is not relevant.
The cyclic cross-correlation between payload and averaged preamble (with phase correction) exhibits an energy peak at the frequency bin of the transmitted symbol $a$ (as illustrated in the figure for $b=a$). 
The RAKE statistic corresponds to the real part of the peak at the selected bin $n=b$ (\textit{i.e.} $\Re\{\tilde{Z}_{ab}[b]\}$). Note that the cyclic cross-correlation followed by the bin selection $n=b$ in the figure is equivalent to the dot-product of $\tilde{\alpha}^*_b(i)$ with the payload left pre-shifted by $b$ as stated in \eqref{eq:Zb_b2}, or as shown in Fig.~\ref{fig:RAKE MF}.

The other main difference with TDEL is that in the TDEL scheme the $\argmax$ operation is just performed after the cyclic correlation to estimate the transmitted symbol whereas for the RAKE, all the candidate symbols need to be tested and the  $\argmax$ operation of $\Re\{\tilde{Z}_{ab}[b]\}$ is performed over each candidate $b$, which increases the computing complexity. We can conclude that the RAKE detector can be seen as a coherent version of TDEL.

\section{simulation results}\label{sec:Simu}

We present in this section several simulation results to evaluate SER performance of our designed receivers.
We consider C1-channel in \eqref{eq:channel_example} plus the following two-path C2-channel:
\begin{equation}
    C_2[k] = \delta[k] + 0.8 \delta[k-5].
    \label{eq:channel_example2}
\end{equation}

We note that simulations are performed with the exact received LoRa signal expression of $\Tilde{r}_a[k]$
because \eqref{eq:ykDech2} is an approximation.
However, theoretical findings are obtained from  \eqref{eq:ykDech2} to derive the new MF and RAKE receivers.

Simulations are performed with the Monte Carlo approach.
A sufficiently high number of trials $N_{\mathrm{trials}}$ is considered.
At each trial, a frame with $N_d = 1000$ data symbols is passed through the channel, the receiver estimates channel parameters when needed and detects these transmitted symbols.
This way, channel parameters are estimated $N_{\mathrm{trials}}$ times and an averaged estimation of the SER performance is then derived.
This prevents statistics bias that may be present when considering a unique channel parameters estimation with higher $N_d$.
Furthermore, simulation figures with ``perfect CSIR" legend (Channel State Information at the Receiver (CSIR)) implies that $\Tilde{\alpha}(i)$ and $k_i$ are perfectly known by the receiver.
Otherwise, the receiver uses the channel parameters scheme presented in Section \ref{sec:ChannelEst} \textcolor{black}{with $k_{\max} = 10$ if need be.}

\subsection{Channel parameters estimation scheme evaluation}

In the next three following figures, we evaluate the channel parameters estimation scheme in terms of 1) the number of required pilot symbols, 2) the $k_i$ estimation behavior (over-, under- and miss-estimation) and 3) the empirical $\rho_p$ threshold value impact on the SER performance.
\subsubsection{number of required pilot symbols}
\begin{figure}[ht]
  \centering
  \includegraphics[width=0.49\textwidth]{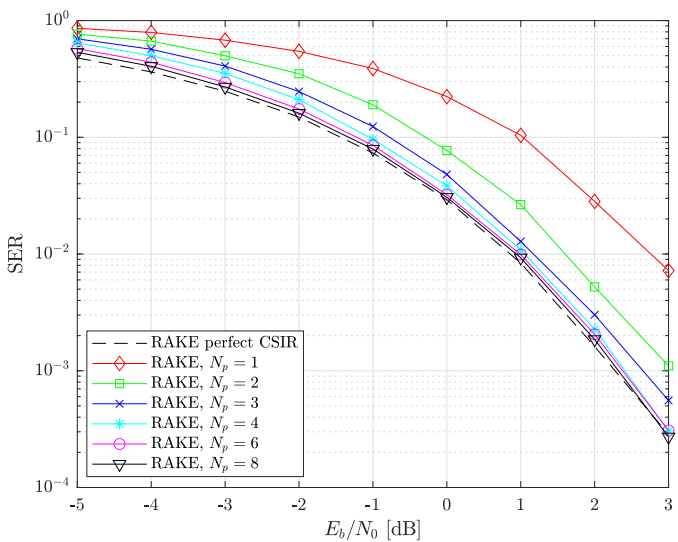}
  \caption{Impact of pilot symbols number $N_p \in \{1,2,3,4,6,8\}$ on RAKE SER performance with C2-channel, $\SF=7$.}
\label{fig:simu_nb_preamble}
\end{figure}

Fig.~\ref{fig:simu_nb_preamble} highlights the benefit of increasing the number of pilot symbols $N_p \in\{1,2,3,4,6,8\}$ required for the channel estimation.
SER performance of RAKE is provided 
with respect to $N_p$ parameter, and compared to RAKE assuming perfect CSIR.
\textcolor{black}{We assume that the number of paths $K$ is known.
That is, the receiver searches the $K-1$ highest magnitudes in the range $n_{k_{max}}$ in \eqref{eqTauEst2} as $\widehat{k}_i$ (for $i=1,\ldots K-1$) and estimate the associated $\Tilde{\alpha}(i)$.}
We consider here C2-channel.
From the figure, we may note that a minimum number of $N_p=4$ pilot symbols is sufficient to be very close to the optimal SER performance with perfect CSIR.
LoRa usually uses $N_p = 8$ pilot symbols, a value giving almost the same SER performance as perfect CSIR, as seen in the figure.
To reduce the complexity and keep very good SER performance, we choose $N_p=6$ value for the rest of simulation results.

\subsubsection{$k_i$ estimation behavior}
\label{subsec:ki estim}
\begin{figure}[ht]
  \centering
  \includegraphics[width=0.49\textwidth]{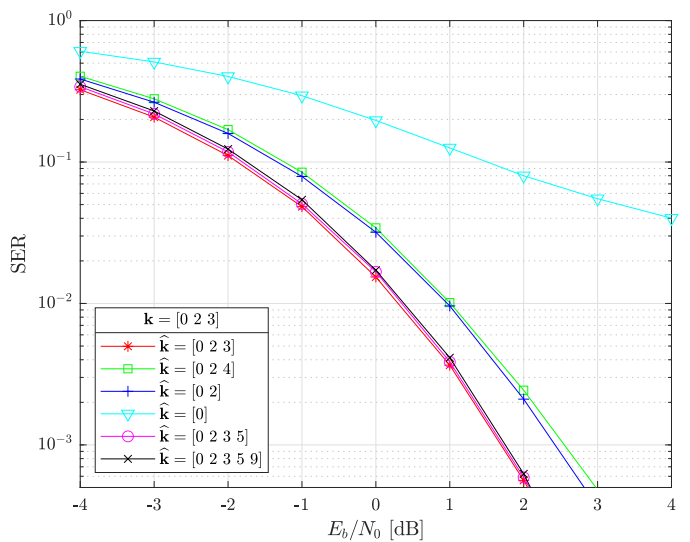}
  \caption{Impact on RAKE SER performance with path delays ($\bm{k}$ over-,  under- and miss-estimation), C1-channel and $\SF=7$.}
\label{fig:test_ki_est_error_impact_C1}
\end{figure}

In Fig.~\ref{fig:test_ki_est_error_impact_C1}, we consider C1-channel having three taps.
We select different sets of $k_i$ values
for the main four situations that can occur: perfect-, under\nobreakseq{-,} over- and miss- estimation of $k_i$.
Note that $|\widehat{\bm{k}}| < K$ and $|\widehat{\bm{k}}| > K$ will be in favor of under and over estimation, respectively.
We assume to be time-synchronized at the first path $k_0=\hat{k}_0=0$.
As seen in the figure, overestimation (i.e. $\widehat{\bm{k}} = [0~2~3~5]$ and $\widehat{\bm{k}} = [0~2~3~5~9]$) does not impact so much SER performance.
In fact, RAKE captures all channel paths energies plus ``ghost paths" pointing to DFT bins having only AWGN.
This is also valid for miss-detection as SER performance between $\widehat{\bm{k}} = [0~2]$ and $\widehat{\bm{k}} = [0~2~4]$ are almost identical.
On the contrary, missing channel paths is very harmful and the benefit of the RAKE receiver is progressively lost as missed paths number grows.
In the figure, the extreme case is for $\widehat{\bm{k}} = [0]$.
RAKE leverages then only the first path leading to the same SER performance as the original coherent receiver (see \eqref{eq:Zb_b2} with $K=1$ and $\tilde{\alpha}(0)=\alpha(0)$).
To prevent this situation, the receiver must then select $\rho_p$ (see \eqref{eqTThreshold}) that reduces channel paths non detection, i.e. $\rho_p$ sufficiently low to lead $|\widehat{\bm{k}}|$ sufficiently high.

\subsubsection{$\rho_p$ threshold impact on SER performance}
\begin{figure}[ht]
  \centering
  \includegraphics[width=0.49\textwidth]{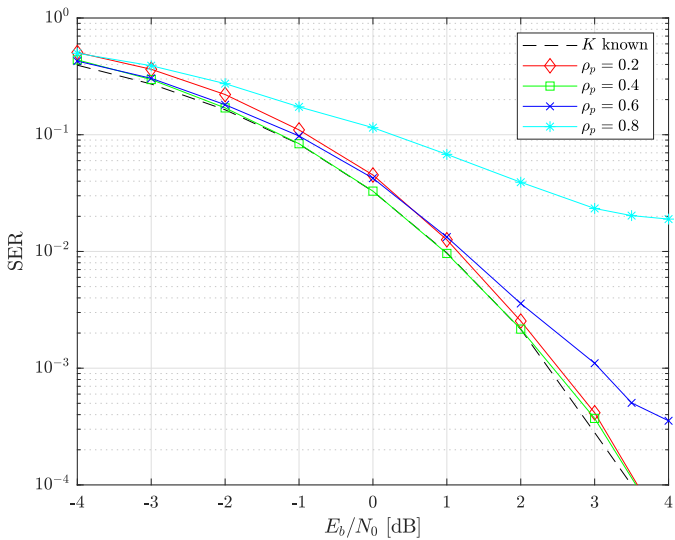}
  \caption{Impact on RAKE SER performance with $\rho_p$ threshold for C2-channel and $\SF=7$.}
\label{fig:test_ki_est_rho_impact_C2}
\end{figure}

The C2-channel is taken as example here.
We set several $\rho_p = \{0.2,0.4,0.6,0.8\}$, activate the channel parameters estimation with $N_p=6$, $k_{\max}=10$ and compute the SER in Fig.~\ref{fig:test_ki_est_rho_impact_C2}.
\textcolor{black}{
We also plot the SER performance assuming the number of paths known (legend: ``$K$ known").
This is the optimal case preventing under- or over-estimation.}
From the figure, we see that the best solution of $\rho_p$ is SNR-dependent.
\textcolor{black}{Overall, $\rho_p = \{0.2,0.4,0.6\}$ perform very well with close SER performance to the $K$ known case, with a small disadvantage for $\rho_p = 0.6$ as the SNR increases.
$\rho_p = \{0.2,0.4\}$ have very similar behavior with a little advantage for $\rho_p = 0.4$ at low SNRs.
$\rho_p = 0.8$ appears to be too high, with too much under-estimation leading to very poor SER performance.
Finally, $\rho_p = 0.4$ seems to be a balanced value for the C2-channel and we keep it for channel parameters estimation with preamble of $N_p=6$ symbols in the rest of the simulations.}


\subsection{RAKE and cand-RAKE SER performance comparison}
\label{subsec:norm Nc}



\begin{figure}[ht]
  \centering
  \includegraphics[width=0.49\textwidth]{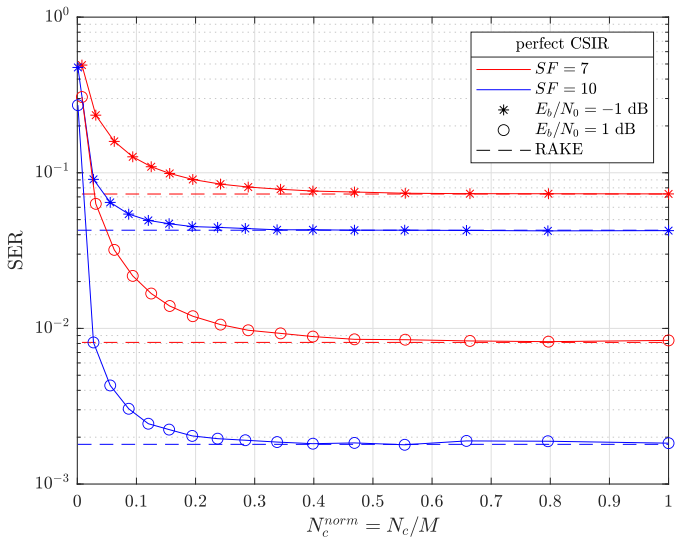}
  \caption{
  SER performance comparison between RAKE and cand-RAKE as a function of the normalized number of candidate symbols $N_c^{\norm} \in~]0,1]$, for
  $\SF = \{7,10\}$, ${E_b/N_0} = \{-1,1\}$~dB, perfect CSIR and C2-channel.}
\label{fig:simu_rake_cand_rake_Nc}
\end{figure}

Fig.~\ref{fig:simu_rake_cand_rake_Nc} compares SER performance of RAKE and cand-RAKE as a function of normalized $N_c$, i.e. $N_c^{\norm} = N_c/M$,
for $\SF = \{7,10\}$ and $E_b/N_0 = \{-1,1\}$ dB.
As we control $N_c$ in the simulations, the first candidate selection method in \ref{subsubsec:fixed_cand_number} is used.
We consider the C2-channel and assume perfect CSIR.
We may see that progressively increasing the number of candidates improves SER performance until converging to the RAKE SER performances with $N_c = M$.
We see that same performances are achieved 
for much lower $N_c$ values than $M$ that are $N_c^{\norm}\approx0.4$ for $\SF=7$ and  $N_c^{\norm}\approx 0.2$ for $\SF=10$, with no significant changes depending on the SNR.

\begin{figure}[ht]
  \centering
    \includegraphics[width=0.49\textwidth]{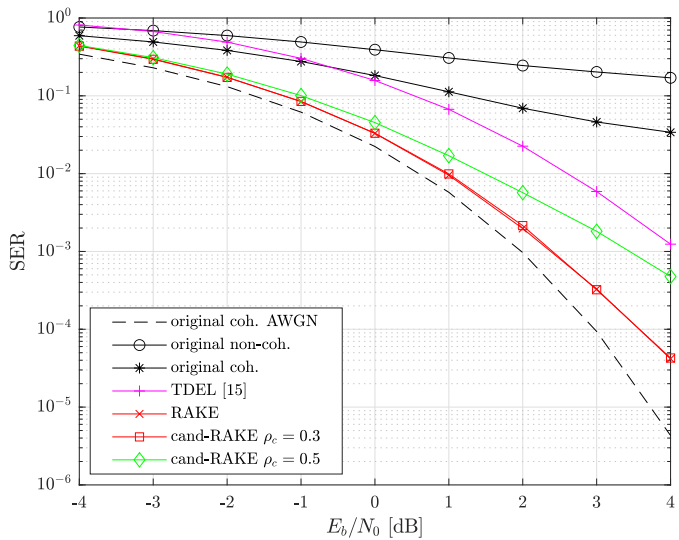}
  \caption{SER performance comparison for C2-channel and $\SF=7$ between RAKE, cand-RAKE, TDEL \cite{guo}, original coherent and non-coherent receivers and the coherent receiver in the AWGN flat-fading channel.
 }
\label{fig:Simu_MF_RAKE_CAND_C1_SF7}
\end{figure}

\begin{figure}[ht]
  \centering
    \includegraphics[width=0.49\textwidth]{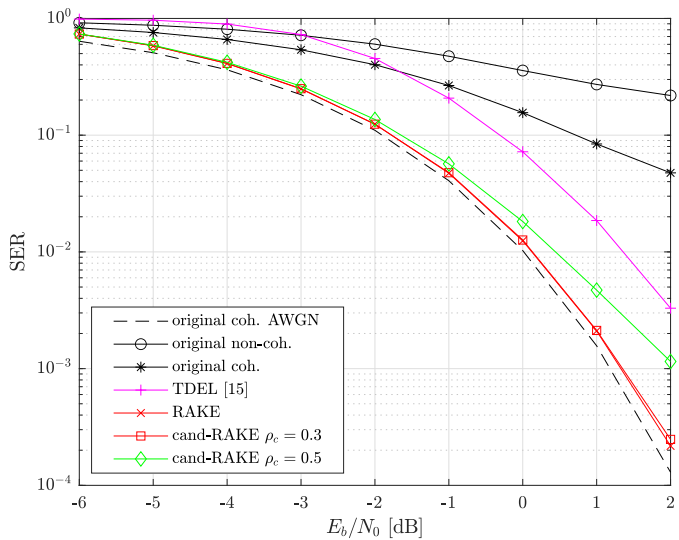}
  \caption{SER performance comparison for C2-channel and $\SF=10$ between RAKE, cand-RAKE, TDEL \cite{guo}, original coherent and non-coherent receivers and the coherent receiver in the AWGN flat-fading channel.
 }
\label{fig:Simu_MF_RAKE_CAND_C1_SF10}
\end{figure}

RAKE and cand-RAKE SER performances are presented in Figs.~\ref{fig:Simu_MF_RAKE_CAND_C1_SF7} and \ref{fig:Simu_MF_RAKE_CAND_C1_SF10}  for $\SF=7$
and $\SF=10$, respectively.
The  variable candidate number method in \ref{subsubsec:variable_cand_number} with $\rho_c = \{0.3,0.5\}$ is now considered.
The corresponding average number of candidates is  denoted as $N_c^{\avg}$ and is reported in Tables \ref{tab:average_cand_SF7} and \ref{tab:average_cand_SF10}, 
for $\SF = 7$ and $\SF = 10$, and for the SNR ranges used in SER plots.
We can clearly deduce a higher computational burden with the lowest threshold $\rho_c=0.3$ in comparison to 
$\rho_c=0.5$ as the average number of candidates has significantly increased. Note that we expect close performance to the RAKE receiver with $\rho_c=0.3$ as
$N_c^{\norm}>0.2$ for $\SF=10$ (whatever the SNR), and for $\SF=7$, $N_c^{\norm}>0.4$ (except at the higher SNR values).
We highlight that although the number of selected candidates for $SF = 7$ and high SNRs is not very high ($N_c^{\norm}<0.4$), there are fewer ``false" candidates due to AWGN, the SER performance of RAKE is then recovered.

\begin{table}[ht]
    \setlength{\tabcolsep}{4pt}
    \centering
\begin{tabular}{c||c|c|c|c|c|c|c|c|c}
$E_b/N_0$ [dB] & -4 & -3 & -2 & -1 & 0 & 1 & 2 & 3 & 4 \\\hline\hline
\hspace{0em}$\rho_c = 0.3$\hfill\mbox{}&&&&&&&&&\\
\hspace{0em}$N_c^{\avg}$\hfill\mbox{} & 76 & 74 & 71 & 66 & 61 & 53 & 45 & 36 & 28 \\
\hspace{0em}$N_c^{\avg}/M$\hfill\mbox{} & 0.59 & 0.58 & 0.55 & 0.52 & 0.48 & 0.41 & 0.35 & 0.28 & 0.22 \\\hline
\hspace{0em}$\rho_c=0.5$\hfill\mbox{}&&&&&&&&&\\
\hspace{0em}$N_c^{\avg}$\hfill\mbox{} & 31 & 30 & 27 & 23 & 19 & 15 & 10 & 7 & 5 \\
\hspace{0em}$N_c^{\avg}/M$\hfill\mbox{} & 0.24 & 0.23 & 0.21 & 0.18 & 0.15 & 0.12 & 0.08 & 0.05 & 0.04
\end{tabular}
\vspace{.1em}
  \caption{Average number and normalized number of selected candidates $N_c^{\avg}$
  (second candidate selection method) as a function of $\rho_c$ and $E_b/N_0$ for $\SF=7$.}
    \label{tab:average_cand_SF7}
\end{table}

\begin{table}[ht]
    \setlength{\tabcolsep}{4pt}
    \centering
\begin{tabular}{c||c|c|c|c|c|c|c|c|c}
$E_b/N_0$ [dB] & -6 & -5 & -4 & -3 & -2 & -1 & 0 & 1 & 2 \\\hline\hline
\hspace{0em}$\rho_c = 0.3$\hfill\mbox{}&&&&&&&&&\\
\hspace{0em}$N_c^{\avg}$\hfill\mbox{} & 517 & 510 & 500 & 483 & 456 & 419 & 364 & 299 & 232 \\
\hspace{0em}$N_c^{\avg}/M$\hfill\mbox{} & 0.50 & 0.49 & 0.49 & 0.47 & 0.45 & 0.41 & 0.36 & 0.29 & 0.23 \\\hline
\hspace{0em}$\rho_c = 0.5$\hfill\mbox{}&&&&&&&&&\\
\hspace{0em}$N_c^{\avg}$\hfill\mbox{} & 159 & 155 & 147 & 136 & 121 & 98 & 72 & 46 & 26 \\
\hspace{0em}$N_c^{\avg}/M$\hfill\mbox{} & 0.16 & 0.115 & 0.14 & 0.13 & 0.12 & 0.10 & 0.07 & 0.04 & 0.03
\end{tabular}
\vspace{.1em}
  \caption{Average number and normalized number of selected candidates $N_c^{\avg}$
  (second candidate selection method) as a function of $\rho_c$ and $E_b/N_0$ for $\SF=10$.}
    \label{tab:average_cand_SF10}
\end{table}

In Figs.~\ref{fig:Simu_MF_RAKE_CAND_C1_SF7} and \ref{fig:Simu_MF_RAKE_CAND_C1_SF10} we have added the original coherent \eqref{eq:DemodLoRaCOH} (labeled ``original coh.") and original non-coherent \eqref{eq:DemodLoRaNCOH} (labeled ``original non-coh.") receivers  for comparisons.
We can see that LoRa legacy demodulation scheme has very poor SER performance bringing to light the need of adapted detection scheme.

We also compare our receiver with the TDEL receiver in \cite{guo} (see Section \ref{subsec:TDEL}).
For our simulations, we set $\rho_{\TDEL} = 0.2$.
As seen in the figures, our scheme outperforms TDEL, especially for $\SF = 10$ where TDEL SER performance is very low at low $E_b/N_0$.
However, at high SNR, TDEL detection scheme presents competitive advantage as SER performance is close to cand-RAKE but with less computational complexity because no search of candidates is necessary and no channel parameters estimation is required.

The cand-RAKE receiver with  $\rho_c=0.3$ has almost identical SER performance as RAKE, for both $\SF = 7$ and $\SF = 10$ and, moreover, a reduced complexity in terms of execution time, as seen in Section \ref{subsubsec:complexity_exec}.
The $\rho_c = 0.5$ value progressively reduces the SER performance as the SNR grows.

Finally, SER performance over the flat-fading AWGN channel (labeled ``original coh. AWGN") of the coherent receiver is presented. 
The same channel energy as  C2-channel is considered  for the flat-fading channel to perform fair comparison.
We conclude that for $\SF=10$, RAKE receiver tends to the optimal original coh. receiver in AWGN.

\section{Conclusion}\label{sec:conclu}

In this paper, we introduced an enhanced LoRa receiver to combat frequency-selective fading MPC.
This receiver coupled with robust channel parameter estimation scheme relies on the simple and elegant MF or equivalent RAKE approaches.
We also developed a candidate variant of these two receivers to reduce complexity.
We came to the conclusion that RAKE is by far preferred to MF in practice as its complexity outperforms MF in all cases.
We point out that we decided to use a simple candidate selection method, although further complexity reduction may be possible by using a threshold that adapts dynamically to the noise level.
The latter needs then to be known by the receiver.
This is an interesting research extension to our work.

When using cand-RAKE, a SER performance-complexity trade-off needs to be taken into consideration.
Table \ref{tab:usage_rec} summarizes the relative complexity of the studied LoRa receivers for MPC. We propose then the following recommendations regarding receivers usage depending on computing capacity of the LoRa system. RAKE achieves the best performance at the price of high computational complexity due to the exhausting search over candidate symbols. However, it can be  implemented on systems that have sufficient computing capacity, typically on the gateway side.
The cand-RAKE with $\rho_c=0.3$ 
is able to achieve performance close to RAKE 
with a significant complexity reduction.
It is well suited for higher-end LoRa transceivers, as for example,
USRP
devices based on Software Defined Radio (SDR).
The cand-RAKE with $\rho_c=0.5$ may be dedicated to mid-end LoRa transceivers with lower computation capabilities.
Finally, we recommend TDEL for low cost LoRa transceivers widely available on the market, with very low computation abilities. However, 
to ensure good performance, TDEL requires
 the SNR value to be large enough.

\begin{table}[ht]
\setlength{\tabcolsep}{2pt}
\begin{center}
\begin{tabular}{l||l}
complexity
& LoRa receiver \\ \hline \hline
high & RAKE ($N_c=M$) \\ \hline
medium  & cand-RAKE ($\rho_c = 0.3$) \\ \hline
low & cand-RAKE ($\rho_c = 0.5$) \\\hline
ultra-low & TDEL \cite{guo} (high SNR)\\
\end{tabular}
\end{center}
\caption{Relative complexity between the LoRa receivers for MPC.}
\label{tab:usage_rec}
\end{table}

We also compared our work to the previous research study on LoRa MPC demodulation scheme TDEL in \cite{guo} and MF/RAKE outperforms TDEL, especially at low $E_b/N_0$ and for high SF.
We recall that MF/RAKE receivers are coherent, that is, channel phase offset and transmitter/receiver desynchronizations must be tackled.
Furthermore, in practice, the channel is non aligned i.e. $k_i$ real valued and the DFT bin energies of channel paths are spread over neighbor bins, making the channel parameters estimation scheme much more difficult to use.
This research may be extended in that sense with this opened challenge.

\section*{Acknowledgment}

This work was jointly supported by the Brest Institute of Computer Science and Mathematics (IBNM) CyberIoT Chair of Excellence of the University of Brest, the Brittany Region and the “Pôle d’Excellence Cyber”.

\bibliographystyle{unsrt}
\bibliography{biblio}

\end{document}